# Diurnal variations of resting-state fMRI data: A graph-based analysis

**Farzad V. Farahani[1*], Waldemar Karwowski[1], Mark D'Esposito[2,3], Richard F. Betzel[4], Magdalena Fafrowicz[5*], Bartosz Bohaterewicz[5], Tadeusz Marek[5], Pamela K. Douglas[6,7]**

[1]Computational Neuroergonomics Laboratory, Department of Industrial Engineering and Management Systems, University of Central Florida, Orlando, FL, USA
[2]Helen Wills Neuroscience Institute, University of California, Berkeley, CA, USA
[3]Department of Psychology, University of California, Berkeley, CA, USA
[4]Department of Psychological and Brain Sciences, Indiana University, Bloomington, IN, USA
[5]Department of Cognitive Neuroscience and Neuroergonomics, Institute of Applied Psychology, Jagiellonian University, Kraków, Poland
[6]Institute for Simulation and Training, University of Central Florida, Orlando, FL, USA
[7]Department of Psychiatry and Biobehavioral Sciences, University of California, Los Angeles, Los Angeles, CA, USA

**\* Correspondence:**
Farzad V. Farahani
farzad.vasheghani@knights.ucf.edu
Magdalena Fafrowicz
magda.fafrowicz@uj.edu.pl

# Abstract

Circadian rhythms (lasting approximately 24 hours) control and entrain a variety of physiological processes ranging from neural activity and hormone secretion to sleep cycles and feeding habits. Despite significant diurnal variation in human brain function, neuroscientists have rarely considered the effects of time-of-day (TOD) on their studies. Moreover, there are inter-individual discrepancies in sleep-wake patterns, diurnal preferences, and daytime alertness (known as chronotypes), which could be associated with human cognition and brain performance. In the present study, we performed graph-theory based network analysis on resting-state functional MRI (rs-fMRI) data to explore the topological differences in whole-brain functional networks between the morning and evening sessions (TOD effect), as well as between extreme morning-type and evening-type participants (chronotype effect). To that end, 62 individuals (31 extreme morning- versus 31 evening-type) underwent two fMRI sessions: about 1 hour after the wake-up time (morning) and about 10 hours after the wake-up time (evening), scheduled in accord with their declared habitual sleep-wake pattern on a regular working day. The findings revealed the significant effect of TOD on the functional connectivity (FC) patterns, but there was no significant difference in chronotype categories. Compared to the morning session, we found relatively increased small-worldness, modularity, assortativity, and synchronization in the evening session, indicating more efficient functional topology. Local measures were changed during the day predominantly across the areas involved in somatomotor, ventral attention, as well as default mode networks. Also, connectivity and hub analyses showed that the somatomotor, ventral attention, and visual networks are the most densely-connected brain areas in both sessions, respectively, with the first being more active in the evening session and the two latter in the morning session. Finally, we performed a correlation analysis to examine whether network properties are associated with the subjective assessments across subjects, most of which happened in the morning session. All these findings contribute to an increased understanding of diurnal fluctuations in resting neural activity and highlight the role of TOD in future brain studies.

# 1. Introduction

Circadian rhythms are natural, internal ~24-hour fluctuations in most living organisms, regulating a variety of physiological functions, including sleep-wake patterns (Borbély, 1982; Dijk and Lockley, 2002; Schmidt et al., 2012), body temperature (Refinetti and Menaker, 1992), endocrine and metabolic rhythms (Hastings et al., 2007), gene expression (Storch et al., 2002), musculoskeletal activity (Aoyama and Shibata, 2017), as well as a wide range of brain functions (Schmidt et al., 2007). Studies of brain function in humans have shown circadian variations in a wide variety of abilities such as attention (Valdez et al., 2005), working memory (Ramírez et al., 2006), motor (Edwards et al., 2007) and visual detection (Tassi et al., 2000). These studies have investigated function at multiple scales of brain organization from that of individual cells and synapses (Gilestro et al., 2009; Kuhn et al., 2016; Vyazovskiy et al., 2008) to brain regions and large-scale FC (Blautzik et al., 2013; Hodkinson et al., 2014; Orban et al., 2020; Shannon et al., 2013; Steel et al., 2019).

A chronotype is a biologically driven circadian typology that generally refers to the individual differences in sleep-wake cycles, diurnal preferences, and alertness throughout the day (Roenneberg et al., 2003; Susman et al., 2007). Traditionally, individuals fall into morning ("early larks") or evening ("night owls") chronotypes. Evening chronotypes typically have phases of behavioral and physiological circadian clocks shifted toward later hours than morning chronotypes (Bailey and Heitkemper, 2001; Kerkhof and Dongen, 1996). Various studies have shown that people with different chronotypes have significantly different diurnal profiles of cognition and behavior (Horne et al., 1980; Norbury, 2020; Schmidt et al., 2007; Valdez et al., 2012). Also, circadian variations in performance-related neural activity have been reported in studies utilizing chronotype-based paradigms (e.g., Facer-Childs et al., 2019; Fafrowicz et al., 2009; Gorfine et al., 2007; Peres et al., 2011; Schmidt et al., 2009, 2012, 2015; Vandewalle et al., 2009, 2011).

However, there have been a limited number of resting-state and task-based functional MRI studies investigating the impact of both TOD and chronotype on behavior, cognition, and neural activity (Blautzik et al., 2013; Cordani et al., 2018; Hodkinson et al., 2014; Jiang et al., 2016; Shannon et al., 2013; Steel et al., 2019; Gorfine and Zisapel, 2009; Marek et al., 2010), often yielding contradictory or sometimes even ambiguous findings. Also, most fMRI studies assume that diurnal fluctuations of brain connectivity patterns as well as human chronotypes are relatively insignificant and are unlikely to lead to a substantial systematic bias into group analysis (Orban et al., 2020).

Here, we examined the effect of TOD on the rs-fMRI data, taking into account subject chronotype, using methodology from network neuroscience. To this end, we directly compared the topological changes (global and local) in FC patterns between morning and evening sessions, as well as the network properties across early larks and night owls. In global analysis, we found a more efficient functional topology as the waking time increased, and local analysis showed significant changes mostly in the somatomotor, ventral attention, and default mode networks. However, we did not find any compelling evidence of the chronotype effect on the network topology itself. Within the context of these findings, it is clear that TOD may influence connectivty patterns in resting state fMRI data, making this variable an potentially important factor to consider in future rs-fMRI experiments.

## 2. Materials and methods

### 2.1. Participants and study procedures

Participants were recruited through online advertisements on the lab's website and Facebook. 5354 volunteers participated in the first stage of selection and were asked to complete two questionnaires: the Chronotype Questionnaire (Oginska et al., 2017) for diurnal preferences assessment and the Epworth Sleepiness Scale (ESS; Johns, 1991) for daytime sleepiness measurements as well as the sleep-wake assessment (real versus ideal wake- and bedtimes). Individuals reporting excessive daytime sleepiness were excluded from the study, as determined by the cut-off points ESS ($\leq 10$ points) questionnaire. 451 participants were divided into morning or evening chronotypes and underwent PER3 VNTR polymorphism genotyping, isolated from buccal swabs using DNA GeneMATRIX Swab-Extract DNA Purification Kit (EURx, Gdańsk, Poland) following manufacturers protocol. Only subjects that were homozygous for the PER3 4 (ES) and PER3 5 (MS) alleles were included for the study. Other selection criteria included an age between 20 and 35 years, right-handed as indicated by the Edinburgh Handedness Inventory (Oldfield, 1971), regular TOD schedule without sleep debt, no neurological or psychiatric disorders, no addiction, normal or corrected-to-normal vision, and no MRI contraindications. Thus, 64 healthy and young participants (39 women, mean age: $23.97 \pm 3.26$ y.o.) were selected for the study that met these criteria.

Resting-state fMRI (rs-fMRI) was performed twice – in the morning (MS) and evening (ES) sessions – about one and ten hours after awakening from a night-time of sleep, respectively. Participants were asked to maintain a regular sleep-wake schedule one week before study, controlled using MotionWatch 8 actigraphs. These actigraphs were also worn during the study days for supervising subjects' sleep length and quality. Furthermore, the night before the morning session, subjects slept in rooms located in the same building where the MRI scanner was located. Individuals abstained from alcohol (48 h) and caffeine (24 h) before MRI scanning sessions and were only allowed to engage in non-strenuous activities during study days. The study was approved by the Institute of Applied Psychology Ethics Committee of the Jagiellonian University. Informed, written consent was provided by all participants in accordance with the Declaration of Helsinki. Demographics, questionnaires and actigraphy results are provided for the morning and evening chronotypes in the Table 1.

Table 1. Demographics, questionnaires and actigraphy results.

| Variables (mean ± SD) | MT (N=31) | ET (N=31) | Sign. |
|---|---|---|---|
| Sex (M/F)[a] | 11/20 | 12/19 | ns |
| Age (years)[b] | 24.45 ± 3.83 | 23.48 ± 2.55 | ns |
| ME[b] | 15.71 ± 2.41 | 28.45 ± 2.39 | * |
| AM[b] | 21.47 ± 3.58 | 22.26 ± 3.51 | ns |
| ESS[b] | 5.52 ± 2.48 | 5.87 ± 3.01 | ns |
| EHI[b] | 86.83 ± 12.92 | 89.19 ± 13.93 | ns |
| VNTR of PER3 | 5/5 | 4/4 | - |
| Declared waketime (hh:mm)[c] | 07:07 ± 62 min | 07:25 ± 48 min | ns |
| Declared bedtime (hh:mm)[c] | 23:24 ± 55 min | 00:06 ± 49 min | * |
| Declared length of perfect sleep (hh:mm)[c] | 08:50 ± 42 min | 08:38 ± 54 min | ns |
| Actigraphy-derived waketime (hh:mm)[c] | 07:43 ± 70 min | 08:16 ± 69 min | ns |
| Actigraphy-derived bedtime (hh:mm)[c] | 23:58 ± 58 min | 00:48 ± 58 min | * |
| Actigraphy-derived length of real sleep (hh:mm)[c] | 07:53 ± 51 min | 07:36 ± 40 min | ns |

Abbreviations: *MT* morning types, *ET* evening types, *ME* morningness/eveningness scale (from Chronotype Questionnaire), *AM* amplitude scale (from Chronotype Questionnaire), *ESS* Epworth Sleepiness Scale, *EHI* Epworth Handedness Inventory, [a] chi-square test, [b] Mann-Whitney U test, [c] Student's t-test, *p*-value < 0.05, *ns* non-significant.

## 2.2. Data acquisition

Magnetic resonance imaging was performed using a 3T Siemens Skyra MR System equipped with a 64-channel head coil. Anatomical images were obtained with the use of sagittal 3D T1-weighted MPRAGE sequence. Ten minutes of resting-state blood-oxygenation-level-dependent (BOLD) images were acquired using a gradient-echo single-shot echo planar imaging sequence with the following parameters: repetition time (TR) = 1,800 ms; echo time (TE) = 27 ms; field of view (FOV) = 256 × 256 mm$^2$; slice thickness = 4 mm; voxel size = 4 × 4 × 4 mm$^3$, with no gap. A total of 34 interleaved transverse slices and 335 volumes acquired in each participants. During scanning, participants were instructed to stay awake and keep their eyes open throughout the scanning session. No other task instruction was provided. Participants' were monitored using an eye tracking system to ensure they were awake throughout the duration of the scan (Eyelink 1000, SR research, Mississauga, ON, Canada).

## 2.3. Data preprocessing

Data preprocessing was performed using DPABI v. 4.2 and SPM 12 both working under Matlab v.2018a (The Mathworks Inc.). Due to the signal equilibration, first 10 time points were discarded. Subsequently, slice timing and realignment with assessment of voxel specific head motion were conducted. The subjects with movements in one or more of the orthogonal directions above 3 mm or rotation above 3° were discarded from the analysis. As the result, a total of four participants were excluded due to the excessive head movements. Then, functional scans were registered using T1 images and then to Montreal Neurological Institute (MNI) space using DARTEL and voxel size of 3 x 3 x 3 mm$^3$. Altogether, seven participants were excluded due to the low quality of the image registrattion. The functional data was spatially smoothed with 4 mm Full Width at Half Maximum (FWHM) kernel. The 24 motion parameters, which were derived from the realignment step were regressed out from the functional data by linear regression as well as five principal components from both cerebrospinal fluid and white matter signals using principal components analysis integrated in a Component Based Noise Correction Method (Behzadi et al., 2007). The global signal was included due to its potential in providing the additional valuable information (Liu et al., 2017) and the signal was band-pass filtered (0.01 – 0.1 Hz).

## 2.4. Brain network construction

A large-scale brain network consists of a finite set of nodes (brain regions) that are connected by links (FC between nodes). In this study, the nodes were specified by parcellation of the whole brain into 7 networks and 200 cortical ROIs from the Schaefer/Yeo atlas (Schaefer et al., 2018). One of the many advantages of this atlas is that each node is preassigned to a system/network (visual, somatomotor, dorsal attention, salience/ventral attention, limbic, frontoparietal, and default mode). The average BOLD time series across all voxels within each ROI were separately extracted. Then, the FC between each pair of ROIs was computed by means of Pearson's correlation coefficient. To improve the normality, the correlation values were converted into z values using Fisher's r-to-z transformation. At this stage, a symmetrical FC matrix (adjacency matrix) with a size of 200×200 was constructed for each subject. Given the controversy over the use of negative correlations (Schwarz and McGonigle, 2011; Wang et al., 2011), we confined the analysis to positive correlations and set the negative coefficients to zero. An overview of our analysis pipeline is shown in **Figure 1**.

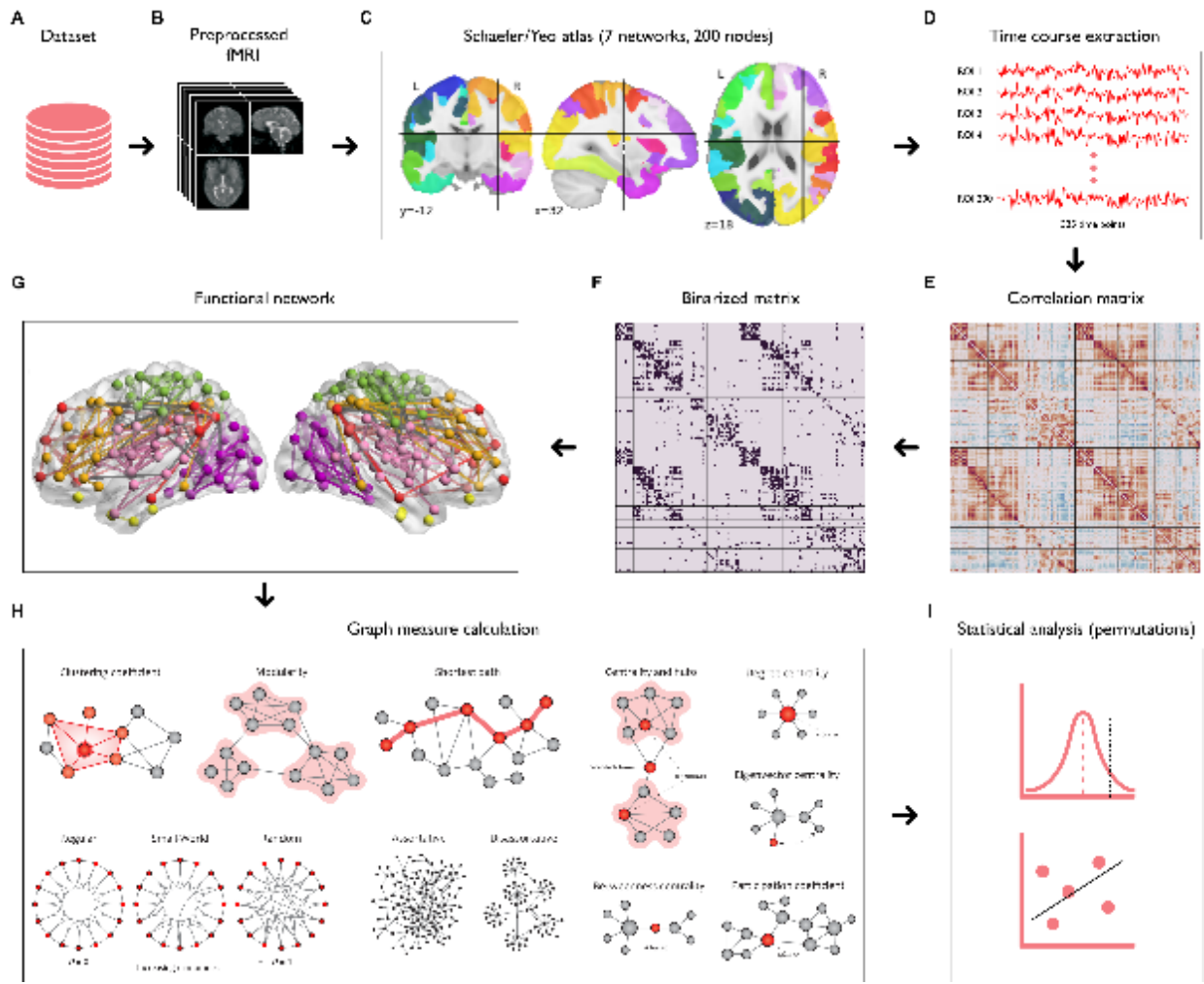

Figure 1. Schematic representation of our graph-based analysis. After preprocessing (B) of the raw rs-fMRI data (A) and parcellating the brain into 200 regions of interest using Schaefer/Yeo atlas (C), corresponding time courses were extracted from each region (D) to compute the correlation matrix (E). To reduce the complexity, the binary correlation matrix (F) and the corresponding functional brain network (G) were constructed, respectively. Then, a set of global and local graph-theoretic measures were derived from the functional network (H). Finally, non-parametric statistics were applied for comparing significant group means and correlations (I).

To exclude the confounding effects of spurious links in interregional connectivity matrices (Power et al., 2011), we adopted a thresholding procedure based on network density to preserve a ratio of the strongest connections and remove weaker connections (van den Heuvel et al., 2017). This procedure leads to equal network density across all subjects (i.e., equal number of edges), which is essential to compare network topology within or between participants (Gamboa et al., 2014). The sparsity threshold we used in this study range from 0.05 to 0.5 with an interval of 0.05, which has been shown to well prevent the formation of disconnected or densely connected networks (Wang et al., 2020). This step is followed by binarizing the thresholded matrices to make the computational complexity more tractable and increase the transparency of network properties (**Figure 2**).

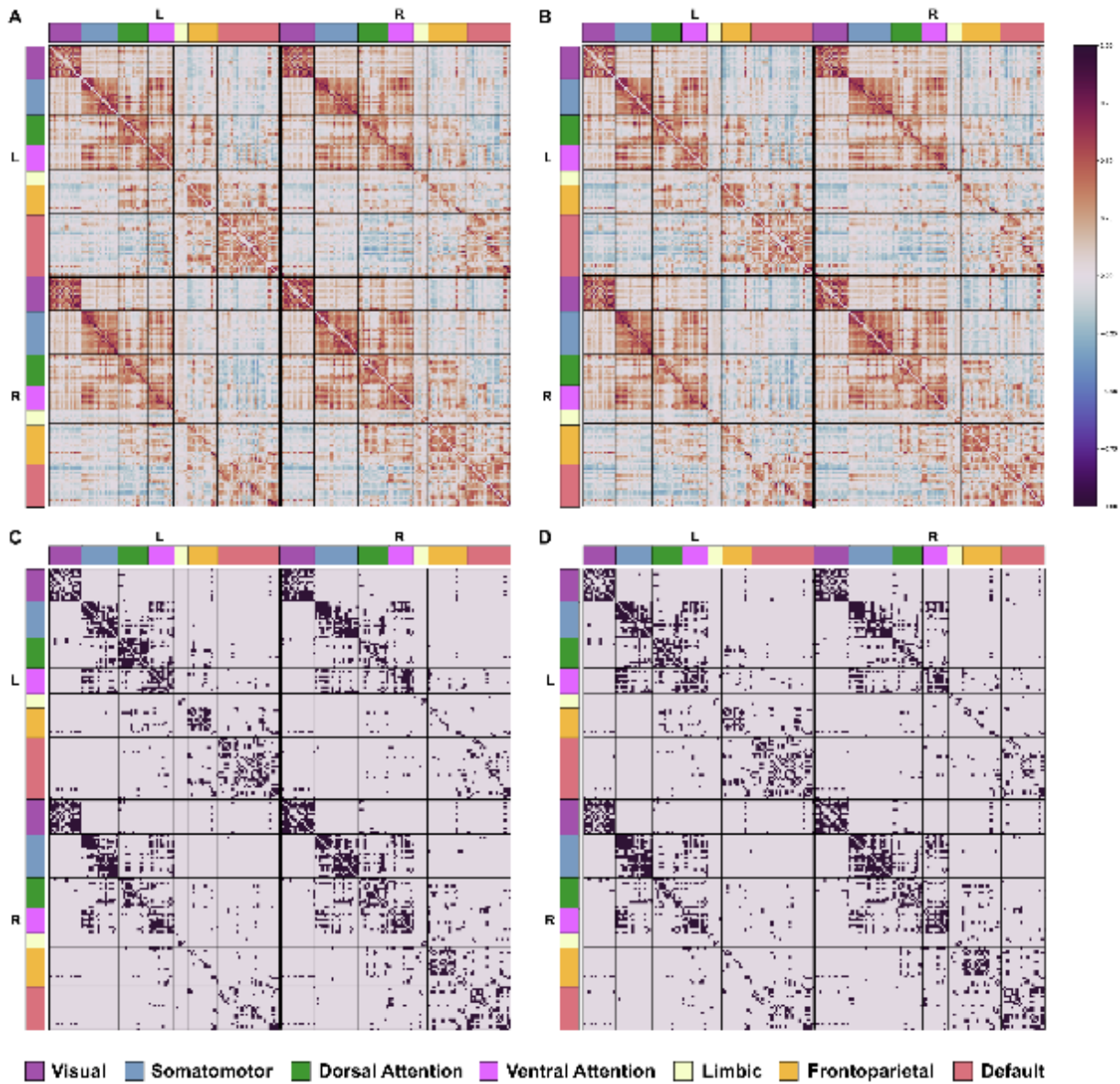

Figure 2. Weighted correlation matrices and binarized connectivity matrices (top 10% strongest connections) for both morning (A and C) and evening (B and D) sessions (averaged across all participants in each session).

## 2.5. Computation of graph metrics

Using binary undirected matrices, we examined the topological properties of functional brain networks for each participant (across a range of cost thresholds) at both global and local levels with the Brain Connectivity Toolbox (BCT[1]; Rubinov and Sporns, 2010). Table 2 provides mathematical definitions and descriptive explanations of each network statistic. Global measures principally describe the functional segregation and integration of brain networks. Thus, we calculated global efficiency, modularity, mean clustering coefficient, characteristic path length, small-worldness, assortativity, and synchronization. Local measures of networks were calculated for each individual node (region) separately, examining the nodal centrality and density of network hubs (i.e., nodes with greater than average number of links). Hubs are generally divided into provincial (which mostly contain local connections within a module) and connector (which contain both local and long-range links to connect nodes in different modules). Thus, we calculated the most common local properties including degree centrality, betweenness centrality, nodal clustering coefficient, nodal efficiency, and participant coefficient (Rubinov and Sporns, 2010).

---
[1] http://www.brain-connectivity-toolbox.net/

Table 2. Mathematical definition and explanation of measures used in this study (for binary and undirected graphs).

| Description | Equation |
|---|---|
| *Degree (nodal)*: It computes the number of nodes neighbors (connections). | For a given node $i$, $$k_i = \sum_{j \in N} a_{ij}$$ $N$: Set of all nodes in the network. |
| *Path length (nodal)*: It quantifies the potential for information transmission and is determined as the average distance from a node to all other nodes. *Characteristic path length (global)*: Average of the path lengths of all other nodes. | Characteristic path length of the network (Watts and Strogatz, 1998), $$L = \frac{1}{n}\sum_{i \in N} L_i = \frac{1}{n}\sum_{i \in N} \frac{\sum_{j \in N, j \neq i} d_{ij}}{n-1}$$ $n$: Number of nodes. $L_i$: Average distance from node $i$ to all other nodes (path length). $d_{ij}$: Shortest path length (distance) between nodes $i$ and $j$. |
| *Efficiency (nodal)*: It computes the efficiency of parallel information transfer of a given node and is determined as the average of the inverse shortest path length from a node to all other nodes. *Efficiency (global)*: Average of the nodal efficiencies of all nodes. | Global efficiency of the network (Latora and Marchiori, 2001), $$E = \frac{1}{n}\sum_{i \in N} E_i = \frac{1}{n}\sum_{i \in N} \frac{\sum_{j \in N, j \neq i} d_{ij}^{-1}}{n-1}$$ $E_i$: Efficiency of node $i$. |
| *Clustering coefficient (nodal)*: The extent to which the neighbors of a given node are interconnected (i.e., fraction of triangles around a node). *Clustering coefficient (global)*: Average of the nodal clustering coefficients of all nodes. To achieve the clustering coefficient of a specific region, one can calculate the mean of the coefficients across that region. | Clustering coefficient of the network (Watts and Strogatz, 1998), $$C = \frac{1}{n}\sum_{i \in N} C_i = \frac{1}{n}\sum_{i \in N} \frac{\sum_{j,k \in N} a_{ij}a_{ik}a_{jk}}{k_i(k_i-1)}$$ $k_i$: Degree of node $i$. $a_{ij}$ is the connection between $i$ and $j$: $a_{ij} = 1$ when link $(i,j)$ exists, and 0 otherwise. There are no self-loops in the network, and therefore $a_{ii} = 0$. |
| *Modularity (global)*: It reflects clusters of densely interconnected nodes with sparse connections among other clusters. | Modularity of the network (Newman, 2004), $$Q = \frac{1}{l}\sum_{i,j \in N}\left(a_{ij} - \frac{k_i k_j}{l}\right)\delta_{m_i,m_j}$$ $L$: Set of all links in the network. $l = \sum_{i,j \in N} a_{ij}$: Number of links. $m_i$: The module containing node $i$, $\delta_{m_i,m_j} = 1$ if $m_i = m_j$, and 0 otherwise (assume that the given network consists of $M$ non-overlapping modules). |
| *Small-worldness (global)*: It is dedicated to graphs in which most nodes are not neighbors but can be reached by any other node with the minimum possible path length. Small-world networks exhibit an intermediate balance between regular and random networks (i.e., they consist of many short-range links alongside a few long-range links), thus reflecting a high clustering coefficient and a short path length. | Small-worldness of the network (Humphries and Gurney, 2008), $$SW = \frac{C_{net}/C_{rand}}{L_{net}/L_{rand}}$$ $C_{net}$ and $L_{net}$ are clustering coefficient and path length of a given network, and $C_{rand}$ and $L_{rand}$ are these measures for an equivalent random network. Small-world networks often have $S \gg 1$. |
| *Assortativity (global)*: The extent to which a network can resist failures in its main components. If $r \geq 0$, the nodes with high degree are more likely to connect to others that are similar in degree (an assortative network), while $r < 0$ reflects that high-degree nodes tend to attach to nodes with low degree (a disassortative network). | Assortativity coefficient of the network (Newman, 2002), $$r = \frac{\frac{1}{l}\sum_{(i,j)\in L} k_i k_j - \left[\frac{1}{l}\sum_{(i,j)\in L}\frac{1}{2}(k_i+k_j)\right]^2}{\frac{1}{l}\sum_{(i,j)\in L}\frac{1}{2}(k_i^2+k_j^2) - \left[\frac{1}{l}\sum_{(i,j)\in L}\frac{1}{2}(k_i+k_j)\right]^2}$$ |
| *Synchronization (global)*: It examines how network nodes fluctuate in the same wave pattern. | Synchronization of the network (Barahona and Pecora, 2002), $$S = \frac{\lambda(2)}{\lambda(M)}$$ $\lambda(2)$: Second smallest eigenvalue of the matrix of $A$. $\lambda(M)$: Largest eigenvalue of the matrix of $A$. $A$: Adjacent matrix of the network. |
| *Betweenness centrality (nodal)*: The ratio of all shortest paths in the network that contain a given node. | Betweenness centrality of node $i$ (Freeman, 1978), $$BC_i = \frac{1}{(n-1)(n-2)} \sum_{\substack{h,j\in N \\ h\neq j, h\neq i, j\neq i}} \frac{\rho_{hj}(i)}{\rho_{hj}}$$ $\rho_{hj}$: Number of shortest paths between $h$ and $j$. $\rho_{hj}(i)$: Number of shortest paths between $h$ and $j$ that use $i$. |
| *Participation coefficient (nodal)*: The distribution of a node's connections across its communities. | Participation coefficient of node $i$ (Guimerà and Nunes Amaral, 2005), $$P_i = 1 - \sum_{m\in M}\left(\frac{k_i(m)}{k_i}\right)^2$$ $M$: Set of non-overlapping modules. $k_i(m)$: Number of links between $i$ and all nodes in module $m$. |

Adopted from: https://sites.google.com/site/bctnet/measures.

### 2.6. Statistical analyses

We applied a non-parametric permutation test (*p*-values were calculated from 50,000 permutations) to investigate the significance of variations and correlations, which does not require any distributional assumptions (Nichols and Holmes, 2002). Also, FDR correction was applied to all statistical tests conducted in this manuscript.

## 3. Results

### 3.1. Global properties

Among the global measures examined, significant differences were found in the small-worldness, network synchronization, and assortativity between morning and evening sessions (**Figure 3**). No compelling evidence of changes was found in other global measures. Small-worldness (**Figure**

**3A**) decreased with higher network sparsity in both sessions. Compared to the morning session, the evening session showed higher small-worldness at sparsity 0.05 and 0.1 ($p < 0.05$, FDR corrected), which did not differ between chronotypes. Assortativity (**Figure 3B**) and network synchronization (**Figure 3C**) increased with higher network sparsity in both sessions. Contrast analysis showed that the assortativity and synchronization were significantly higher during the evening session than the morning session at sparsity 0.3 to 0.5 ($p < 0.05$, FDR corrected), which did not differ between chronotypes.

There were also significant differences in the modularity index between the two sessions at lower sparsity ($p < 0.05$, FDR corrected; **Figure 4**). Modularity was higher in the evening session as compared to the morning session.

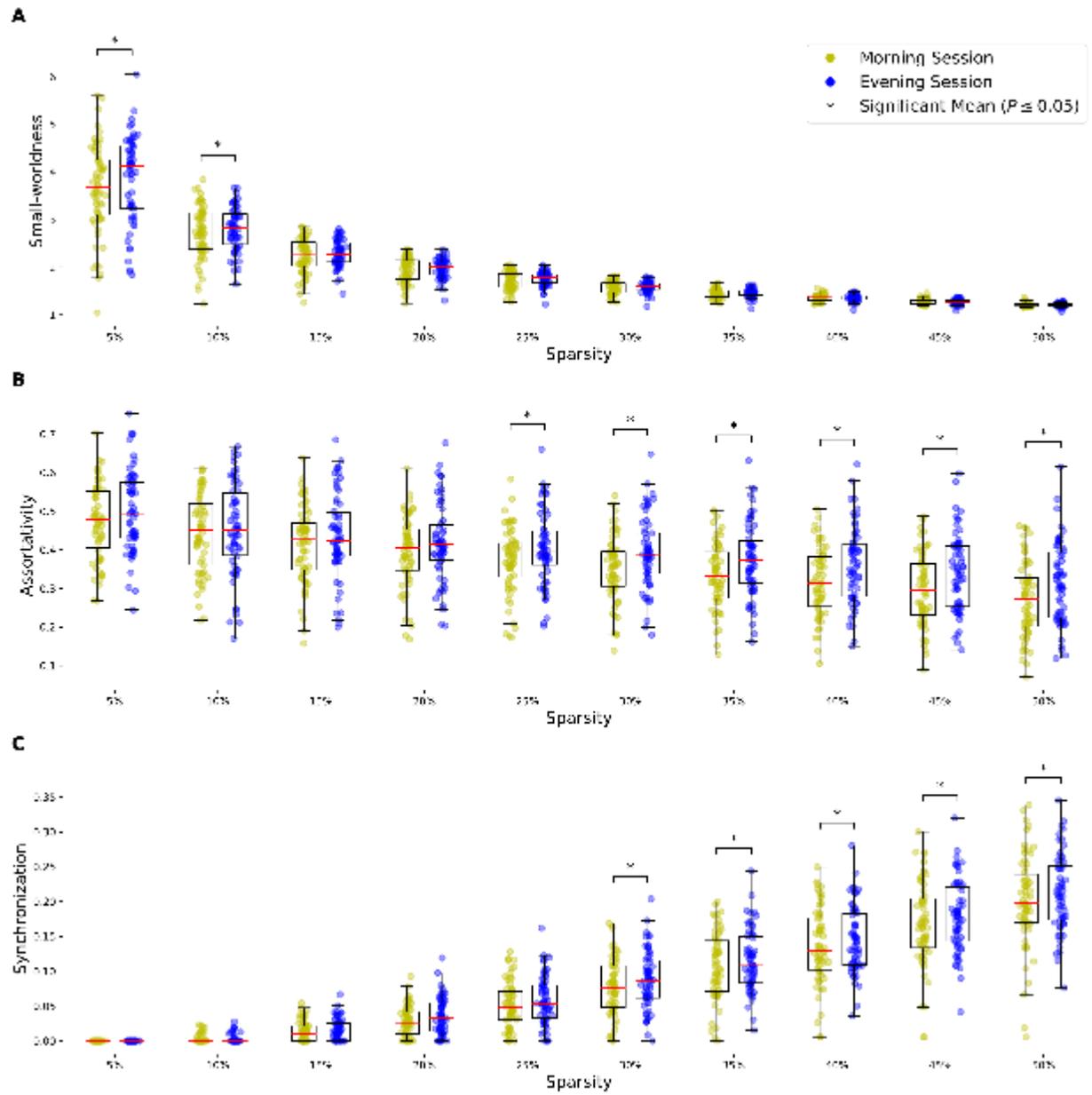

Figure 3. Differences of the small-worldness (A), assortativity (B) and synchronization (C) between the morning and evening sessions at the threshold values of 0.05 to 0.5 (*p*-values were computed using 50,000 permutations).

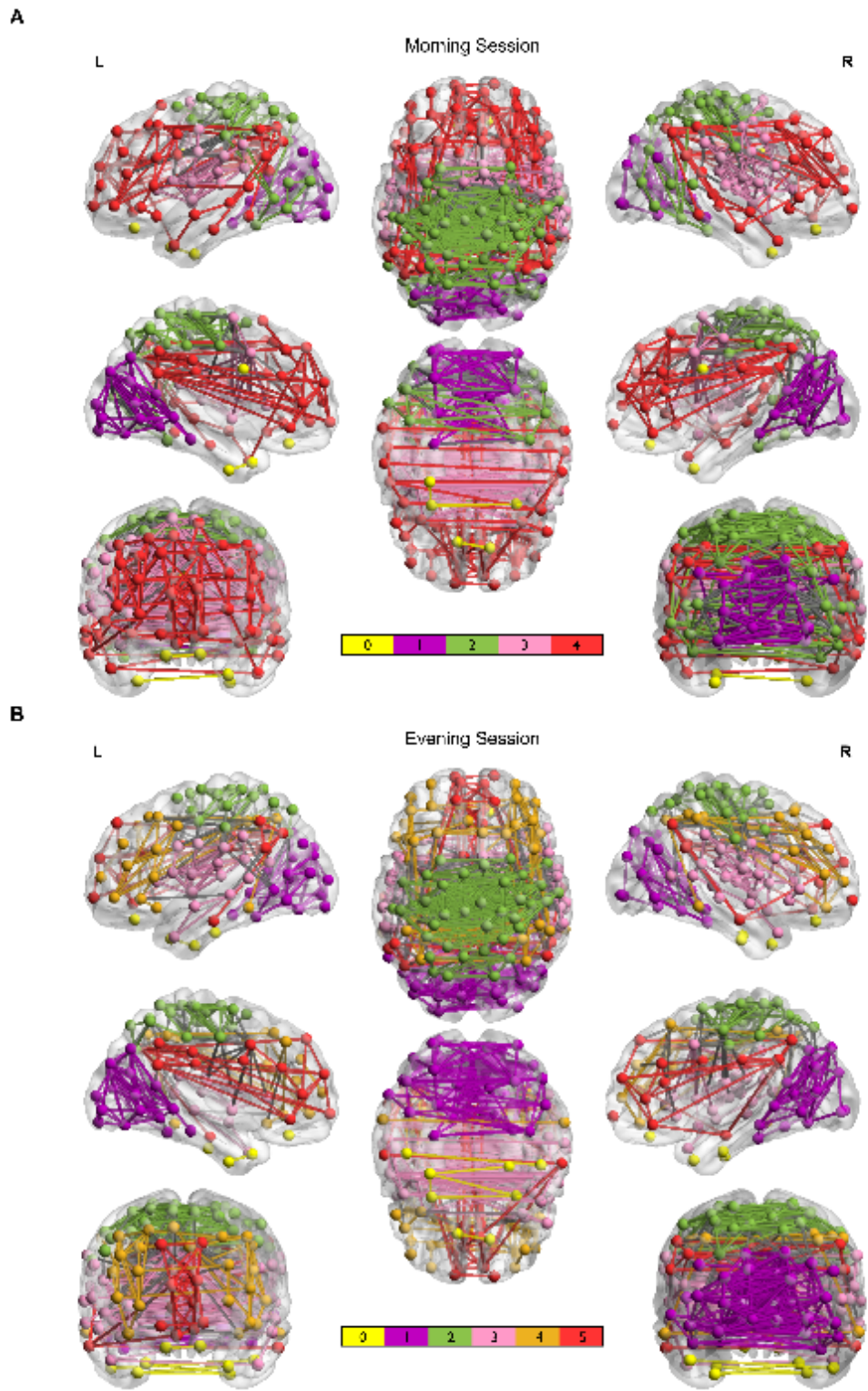

Figure 4. Modular brain network of the morning session and the evening session at a sparsity of 0.05 (visualization is based on the group-averaged data, while statistical analyses were carried out using individual subjects). There were four modules in the morning session (A) and five modules in the evening session (B). The horizontal color bar indicates the color coding of the modules (non-assigned nodes are marked with zero). Node connections within each module are represented in the module's color, while connections between different modules are represented in gray.

## 3.2. Local properties

**Table 3** summarizes brain regions that significantly changed throughout the day using a set of nodal/local properties. The areas under the curve (AUC) were calculated for each network measure to provide a scalar that does not depend on a specific threshold selection (Wang et al., 2009; Zhang et al., 2011). As presented in **Table 3**, we found significant differences between the morning and evening sessions in terms of their degree centrality, betweenness centrality, clustering coefficient, and nodal efficiency. Most of these differences involved regions and their homotopic partners in the opposite hemisphere. No significant differences in participation coefficient and nodal shortest path were found between the two sessions ($p > 0.05$).

The AUC analysis for degree centrality, betweenness centrality, clustering coefficient, and nodal efficiency for all 200 brain regions are presented in **Figure 5**. Compared with the morning session, the evening session showed significantly increased degree centrality in the left somatomotor network and areas such as the superior temporal gyrus and postcentral gyrus, and decreased degree centrality in the left ventral attention and control network and areas such as supramarginal, middle frontal, and angular gyri ($p < 0.05$, FDR corrected). Similar results to degree centrality were obtained for the nodal efficiency. Betweenness centrality analysis also showed a significant increase in the evening session compared to the morning session in areas such as the bilateral precuneous, right postcentral gyrus, and right superior parietal gyrus. Finally, nodal clustering coefficient was increased in the evening session as compared to the morning session in the left dorsal anterior cingulate gyrus and right temporal gyrus (middle and superior), while it was decreased in the right angular gyrus over the course of the day.

Table 3. List of brain ROIs that significantly changed throughout the day (*p*-values were computed using 50,000 permutations followed by FDR correction).

| ROI | Schaefer node label | Cortical areas | MNI coordinates | | | Adjusted *p*-value | | | |
|---|---|---|---|---|---|---|---|---|---|
| | | | x | y | z | Degree Centrality | Betweenness Centrality | Clustering Coefficient | Nodal Efficiency |
| 15 | LH_SomMot_1 | L_Superior temporal gyrus | -51 | -4 | -2 | 0.0053 | | | 0.0069 |
| 16 | LH_SomMot_2 | L_Superior temporal gyrus | -53 | -24 | 9 | 0.0132 | | | 0.0164 |
| 25 | LH_SomMot_11 | L_Superior parietal gyrus | -31 | -46 | 63 | 0.0287 | | | 0.0234 |
| 30 | LH_SomMot_16 | L_Postcentral gyrus | -19 | -31 | 68 | 0.0056 | | | 0.0075 |
| 46 | LH_SalVentAttn_ParOper_3 | L_Supramarginal gyrus | -60 | -39 | 36 | 0.0250 | | | 0.0297 |
| 51 | LH_SalVentAttn_PFCl_1 | L_Middle frontal gyrus (dorsal prefrontal cortex) | -28 | 43 | 31 | 0.0096 | | | 0.0053 |
| 63 | LH_Cont_Par_3 | L_Angular gyrus | -45 | -42 | 46 | 0.0071 | | | 0.0092 |
| 73 | LH_Cont_Cing_2 | L_Dorsal anterior cingulate gyrus | -3 | 4 | 30 | | | 0.0090 | |
| 94 | LH_Default_PFC_12 | L_Superior frontal gyrus (posterior segment) | -24 | 25 | 49 | | | 0.0283 | |
| 99 | LH_Default_pCunPCC_4 | L_Precuneous/Posterior Cingulate Cortex | -6 | -54 | 42 | | 0.0253 | | |
| 117 | RH_SomMot_2 | R_Superior temporal gyrus | 64 | -23 | 8 | 0.0215 | | | |
| 122 | RH_SomMot_7 | R_Postcentral gyrus | 58 | -5 | 31 | | 0.0237 | | |
| 138 | RH_DorsAttn_Post_4 | R_Angular gyrus | 46 | -38 | 49 | | | 0.0055 | |
| 139 | RH_DorsAttn_Post_5 | R_Superior parietal gyrus | 41 | -31 | 46 | | 0.0111 | | |
| 187 | RH_Default_Temp_3 | R_Superior temporal gyrus | 55 | -6 | -10 | | | 0.0174 | |
| 188 | RH_Default_Temp_4 | R_Middle temporal gyrus | 63 | -27 | -6 | | | 0.0103 | |
| 200 | RH_Default_pCunPCC_3 | R_Precuneous/Posterior Cingulate Cortex | 6 | -58 | 44 | | 0.0125 | | |

Abbreviations: *MNI* Montreal Neurological Institute space; *LH* left hemisphere; *RH* right hemisphere; *SomMot_1*, *SomMot_2*, *SomMot_7*, *SomMot_11*, *SomMot_16*, first, second, seventh, eleventh, and sixteenth segment of the Somatomotor Network parcel; *DorsAttn_Post_4*, *DorsAttn_Post_5*, fourth and fifth segment of the Posterior Dorsal Attentional Network parcel; *SalVentAttn_ParOper_3*, third segment of the Parietal Operculum Salience/Ventral Attention Network parcel; *SalVentAttn_PFCl_1*, first segment of the Lateral Prefrontal Cortex Salience/Ventral Attention Network parcel; *Cont_Par_3*, third segment of the Parietal Control Network parcel; *Cont_Cing_2*, second segment of the Cingulate Control Network parcel; *Default_Temp_3*, *Default_Temp_4*, third and fourth segment of the Temporal Default Network parcel; *Default_PFC_12*, twelfth segment of the Prefrontal Cortex Default Network parcel; Default_pCunPCC_3, Default_pCunPCC_4, third and fourth segment of the Precuneus Posterior Cingulate Cortex Default Network parcel.

Note: Labels from the Yeo and Schaefer Atlas, available from: https://github.com/ThomasYeoLab/CBIG/blob/master/stable_projects/brain_parcellation/Schaefer2018_LocalGlobal/Parcellations/MNI/Centroid_coordinates/Schaefer2018_200Parcels_7Networks_order_FSLMNI152_1mm.Centroid_RAS.csv.

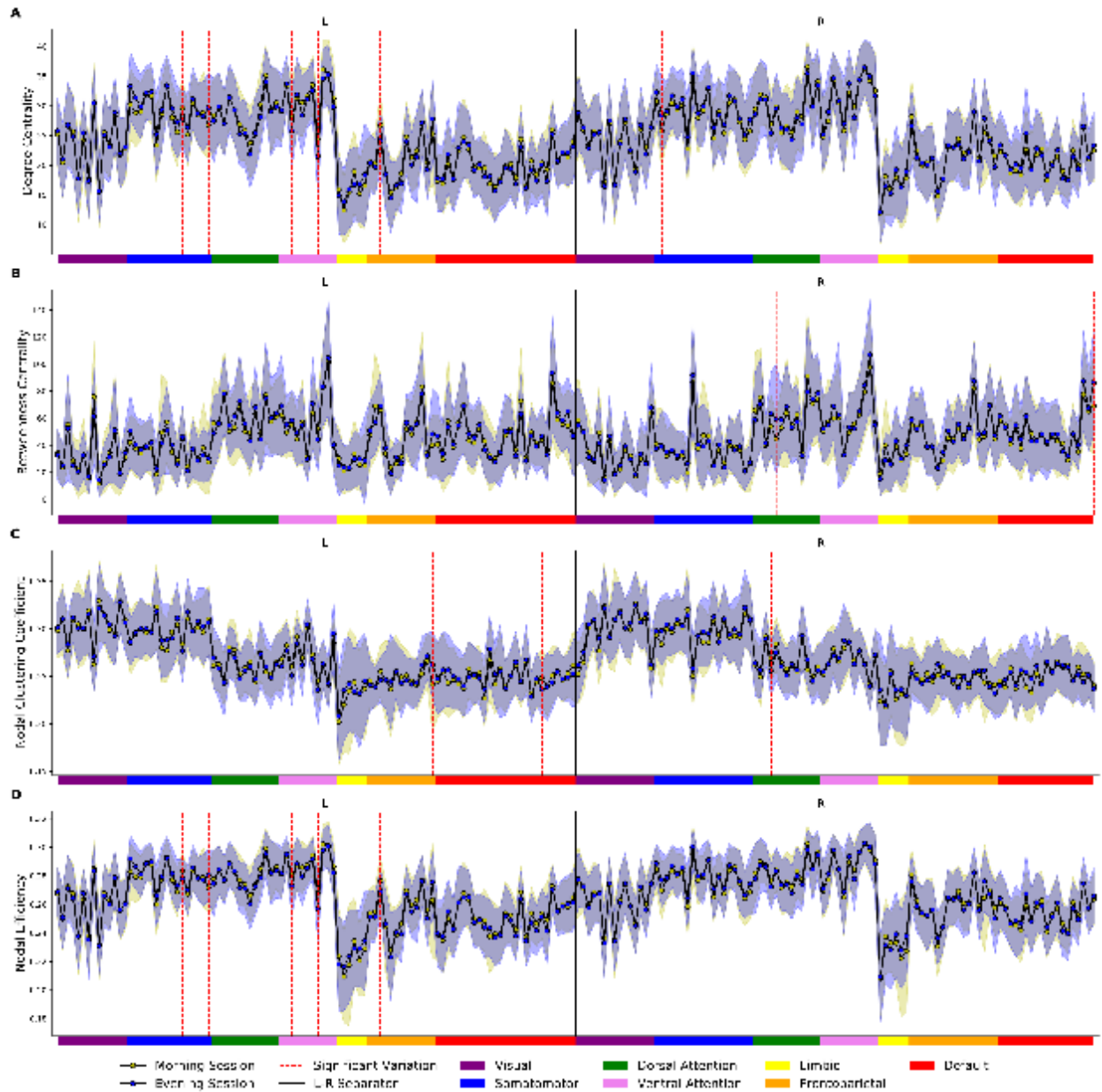

Figure 5. Area under the curve in the morning and evening sessions for degree centrality (A), betweenness centrality (B), clustering coefficient (C), and nodal efficiency (D) of all 200 ROIs. Each node (either in the left or right hemisphere) is labeled by a color matched to Schaefer/Yeo 7 network parcellation. Significant diurnal changes are represented by dashed red lines.

3.3. Hub analysis

In this subsection, using a pre-determined modular classification, which includes the visual, somatomotor, dorsal attention, ventral attention, limbic, frontoparietal, and default mode network (Yeo et al., 2011), we identified network hubs along with their types (i.e., connector or provincial) in both morning and evening sessions (**Table 4**). The results are based on the mean connectivity matrix (across all participants for each corresponding session), and for simplicity in visualizing them later, we chose network density of 0.05. As presented in **Table 4**, the identified hubs are almost overlapping in the morning and evening sessions, except for minor changes in areas such as right SomMot_3 (posterior insula), right Post_5 (superior parietal gyrus), right PrCv_1 (precentral gyrus), and left pCun_1 (cuneus). We found that approximately 25% of somatomotor network nodes play a hub role in both morning and evening sessions, making this module the densest part of the brain during resting state. Notably, somatomotor hubs are either provincial (i.e., within modular connections) or connector (i.e., between modular connections), while hubs in the ventral attention network are connector and hubs in the visual network are provincial.

To visualize the results, connectograms of both sessions are illustrated in **Figure 6** using Circos software (Krzywinski et al., 2009). Parcelled elements on the outermost circle indicate the 200 Schaefer/Yeo brain areas marked with a unique RGB code and belong to one of the predefined modules in each hemisphere. This outer circle circumscribes five inner circular heatmaps created to represent the values of five distinct centrality measures. The range for each of these measures is from the minimum to maximum assumed value. Toward the center, these measures are degree centrality, participation coefficient, K-coreness centrality, eigenvector centrality, and PageRank. The values of all measures, as well as the functional connections in each of the connectograms, are derived from the average of all individuals in the corresponding session. The red and black curves show the functional connections between and within modules, respectively. An unambiguous abbreviation scheme was created to label each parcellation, as summarized in Appendix Table A1.

Table 4. Identifying hubs in each brain network for both sessions (at a sparsity of 0.05).

| Networks | L/R | Nodes | Session | Hubs (P, C) | Hub Regions $^{(Type)}$ |
|---|---|---|---|---|---|
| **Visual** | L | 14 | M | 3 (3, 0) | Vis_1$^{(P)}$, Vis_13$^{(P)}$, Vis_14$^{(P)}$ |
| | | | E | 3 (3, 0) | Vis_1$^{(P)}$, Vis_13$^{(P)}$, Vis_14$^{(P)}$ |
| | R | 15 | M | 3 (3, 0) | Vis_3$^{(P)}$, Vis_4$^{(P)}$, Vis_12$^{(P)}$ |
| | | | E | 3 (3, 0) | Vis_3$^{(P)}$, Vis_4$^{(P)}$, Vis_12$^{(P)}$ |
| **Somatomotor** | L | 16 | M | 4 (2, 2) | SomMot_4$^{(C)}$, SomMot_5$^{(C)}$, SomMot_8$^{(P)}$, SomMot_13$^{(P)}$ |
| | | | E | 4 (2, 2) | SomMot_4$^{(C)}$, SomMot_5$^{(C)}$, SomMot_8$^{(P)}$, SomMot_13$^{(C)}$ |
| | R | 19 | M | 5 (3, 2) | SomMot_5$^{(C)}$, SomMot_6$^{(C)}$, SomMot_11$^{(P)}$, SomMot_15$^{(P)}$, SomMot_16$^{(P)}$ |
| | | | E | 6 (4, 2) | **SomMot_3$^{(P)}$**, SomMot_5$^{(C)}$, SomMot_6$^{(C)}$, SomMot_11$^{(P)}$, SomMot_15$^{(P)}$, SomMot_16$^{(P)}$ |
| **Dorsal Attention** | L | 13 | M | 0 (0, 0) | — |
| | | | E | 0 (0, 0) | — |
| | R | 13 | M | 2 (0, 2) | **Post_5$^{(C)}$, PrCv_1$^{(C)}$** |
| | | | E | 0 (0, 0) | — |
| **Ventral Attention** | L | 11 | M | 5 (0, 5) | ParOper_2$^{(C)}$, FrOper_1$^{(C)}$, FrOper_3$^{(C)}$, FrOper_4$^{(C)}$, Med_1$^{(C)}$ |
| | | | E | 5 (0, 5) | ParOper_2$^{(C)}$, FrOper_1$^{(C)}$, FrOper_3$^{(C)}$, FrOper_4$^{(C)}$, Med_1$^{(C)}$ |
| | R | 11 | M | 4 (0, 4) | TempOccPar_3$^{(C)}$, FrOper_2$^{(C)}$, FrOper_4$^{(C)}$, Med_1$^{(C)}$ |
| | | | E | 4 (0, 4) | TempOccPar_3$^{(C)}$, FrOper_2$^{(C)}$, FrOper_4$^{(C)}$, Med_1$^{(C)}$ |
| **Limbic** | L | 6 | M | 0 (0, 0) | — |
| | | | E | 0 (0, 0) | — |
| | R | 6 | M | 0 (0, 0) | — |
| | | | E | 0 (0, 0) | — |

| | L/R | | M/E | | |
|---|---|---|---|---|---|
| **Frontoparietal** | L | 13 | M | 0 (0, 0) | — |
| | | | E | 1 (0, 1) | **pCun_1**[(C)] |
| | R | 17 | M | 1 (0, 1) | pCun_1[(C)] |
| | | | E | 1 (0, 1) | pCun_1[(C)] |
| **Default Mode** | L | 27 | M | 0 (1, 0) | **pCunPCC_2**[(P)] |
| | | | E | 0 (1, 0) | |
| | R | 19 | M | 0 (0, 1) | **pCunPCC_1**[(C)] |
| | | | E | 0 (0, 1) | |

Abbreviations: *L/R* left or right hemisphere; *M* morning; *E* evening session; *P* provincial; *C* connector.
Note: Regions information is provided in the Appendix Table A1.

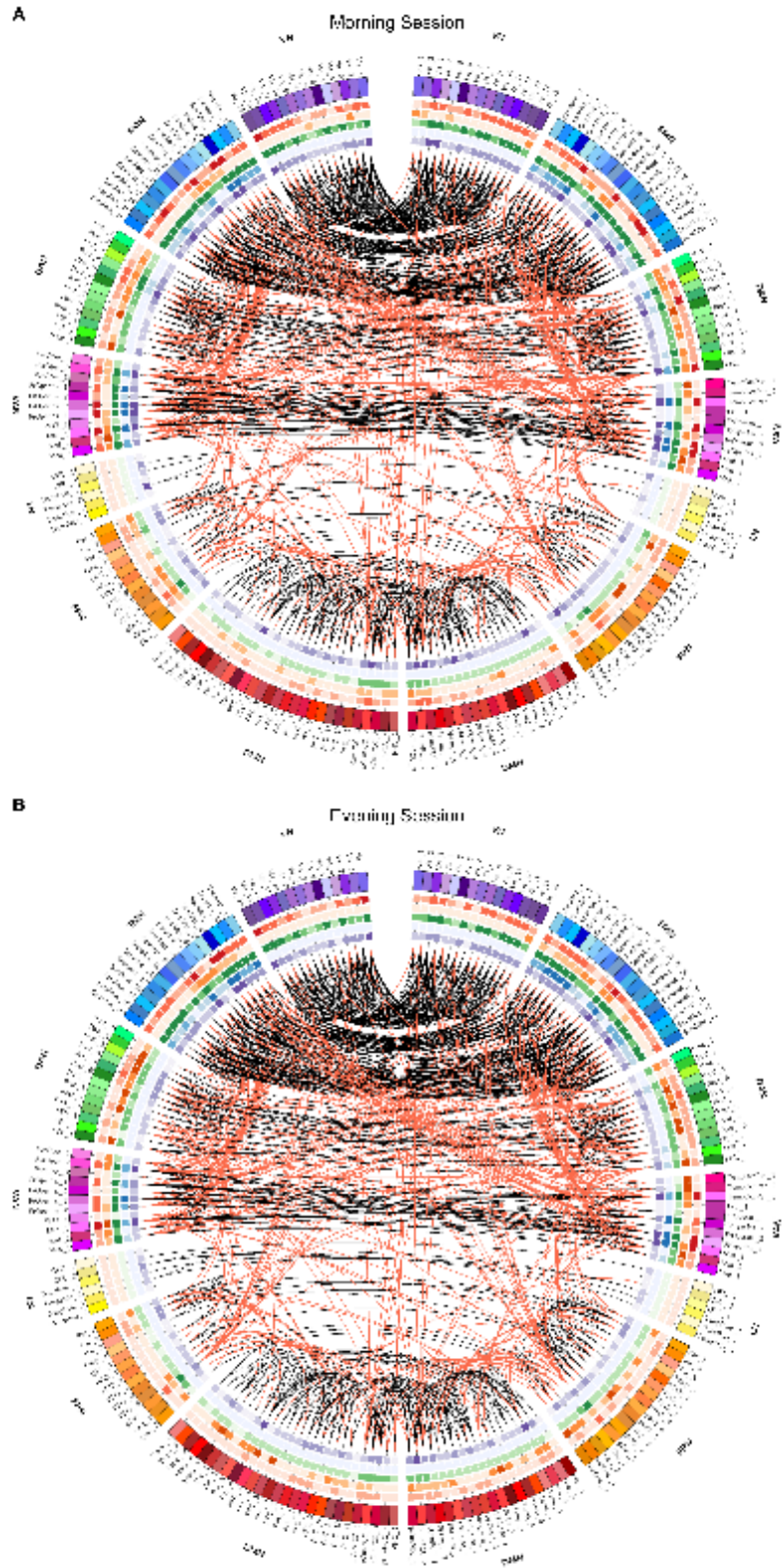

Figure 6. The mean connectogram across all participants in both morning (A) vs. evening (B) conditions at the thresholding value of 0.05. Abbreviations: *VN* visual network, *SMN* somatomotor network, *DAN* dorsal attention network, *VAN* ventral attention network, *LN* limbic network, *FPN* frontoparietal network, *DMN* default mode network.

## 3.4. Correlation analysis

A correlation analysis was performed to investigate whether global and nodal measures are significantly correlated with the variable of interest (e.g., ME scale, AM scale, ESS, or any other cognitive/behavioral variables) across subjects while controlling for the differences of the covariates of no interest (e.g., age, gender, and clinical variables). ME, AM, and ESS scores determine "person's chronotype preference", "how strong the preference is", and "sleepiness during the day", respectively. Overall, the number of significant associations was greater in the morning session rather than the evening session. From global perspective (**Table 5** and **Figure 7**), correlation analysis revealed significant negative associations between AM score and small-worldness and modularity in the morning session ($p < 0.05$, FDR corrected). We also found significant positive relations between ESS and average path length and assortativity in the morning session, as well as positive correlations between AM score and path length and assortativity in the evening session.

From a nodal perspective, we found significant correlations between degree centrality of various parts of the brain in both hemispheres and the subjective indicators (e.g., ME scale, AM scale, ESS), mostly across the morning scanning session (**Table 6** and **Figure 8**). In the morning session, there were significant negative correlations between: AM score and areas within the default network including left rostral anterior cingulate gyrus (Default_PFC_6), left precuneus (Default_PCC_2 and Default_PCC_4), right medial prefrontal cortex (Default_PFCm_4) and right posterior cingulate cortex (Default_PCC_2); ESS and left pole of superior temporal gyrus (Limbic_TempPole_3), right lateral fronto-orbital gyrus (Cont_PFCl_1) and right pole of middle temporal gyrus (Default_Temp_1); ME score and left postcentral gyrus (SomMot_4), left pole of superior temporal gyrus (Limbic_TempPole_4), as well as positive associations between: AM score and bilateral precentral gyrus (left DorsAttn_FEF_1 and right SomMot_11); ME score and bilateral insular (SalVentAttn_FrOper_2 and Cont_PFCv_1), left anterior cingulate gyrus (Default_PFC_8), and right precentral gyrus (DorsAttn_FEF_1).

During the evening session, there were significant positive associations between: AM score and left middle occipital gyrus (Vis_11), right fusiform gyrus (Vis_2) and right superior occipital gyrus (Vis_14), as well as negative correlations between: AM score and left precentral gyrus (DorsAttn_PrCv_1), bilateral lateral fronto-orbital gyrus (Limbic_OFC_1 and Limbic_OFC_2), and right middle temporal gyrus (Default_Temp_4). No significant correlation between ESS and ME scores and degree centrality of brain regions was found.

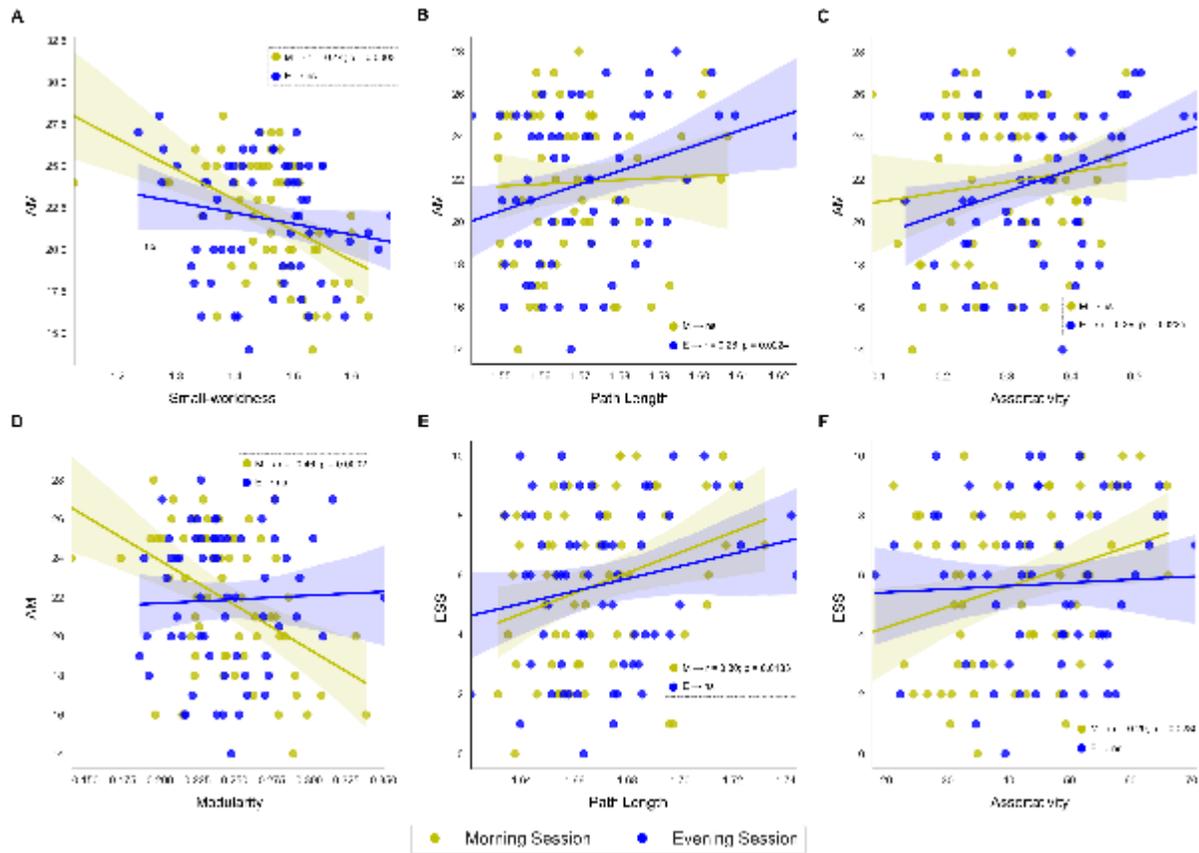

Figure 7. Significant associations between the global measures and questionnaire variables (AM, ESS and ME scores). AM and Small-worldness (A), AM and path length (B), AM and assortativity (C), AM and modularity (D), ESS and path length (E), and ESS and assortativity (D). Yellow and blue colors represent the scatterplots for morning and evening session, respectively. In each panel, one of the correlations related to the morning or evening session is statistically significant and the other is not-significant.

Table 5. Global analysis: significant correlations between global metrics and AM, ESS and ME scores (n = 62) for morning and evening sessions. The significance level is set at a $p < 0.05$ with non-parametric permutations.

|  | Correlation (adjusted $p$-value) | | | | | |
|---|---|---|---|---|---|---|
|  | Morning session | | | Evening session | | |
|  | AM | ESS | ME | AM | ESS | ME |
| Small-worldness | -0.44 (0.0003) | — | — | — | — | — |
| Path Length | — | 0.30 (0.0183) | — | 0.28 (0.0224) | — | — |
| Modularity | -0.46 (0.0002) | — | — | — | — | — |
| Assortativity | — | 0.29 (0.0230) | — | 0.29 (0.0225) | — | — |

Table 6. Nodal analysis: significant correlations between degree centrality and AM, ESS and ME scores (n = 62) for morning and evening sessions. The significance level is set at a $p < 0.05$ with non-parametric permutations.

| | | ROI (Schaefer/Yeo Atlas) | Correlation (adjusted $p$-value) | | | | | |
|---|---|---|---|---|---|---|---|---|
| | | | Morning Session | | | Evening Session | | |
| | | | AM | ESS | ME | AM | ESS | ME |
| Left Hemisphere | 11 | Vis_11 | | | | 0.30 (0.0189) | | |
| | 18 | SomMot_4 | | | -0.37 (0.0029) | | | |
| | 41 | DorsAttn_FEF_1 | 0.38 (0.0023) | | | | | |
| | 43 | DorsAttn_PrCv_1 | | | | -0.33 (0.0091) | | |
| | 48 | SalVentAttn_FrOper_2 | | | 0.37 (0.0024) | | | |
| | 55 | Limbic_OFC_1 | | | | -0.32 (0.0149) | | |
| | 59 | Limbic_TempPole_3 | | -0.42 (0.0007) | | | | |
| | 60 | Limbic_TempPole_4 | | | -0.33 (0.0076) | | | |
| | 88 | Default_PFC_6 | -0.32 (0.0113) | | | | | |
| | 90 | Default_PFC_8 | | | 0.30 (0.0200) | | | |
| | 97 | Default_PCC_2 | -0.39 (0.0015) | | | | | |
| | 99 | Default_PCC_4 | -0.32 (0.0093) | | | | | |
| Right Hemisphere | 102 | Vis_2 | | | | 0.30 (0.0191) | | |
| | 114 | Vis_14 | | | | 0.31 (0.0159) | | |
| | 126 | SomMot_11 | 0.30 (0.0172) | | | | | |
| | 145 | DorsAttn_FEF_1 | | | 0.33 (0.0080) | | | |
| | 160 | Limbic_OFC_2 | | | | -0.32 (0.0116) | | |
| | 169 | Cont_PFCv_1 | | | 0.31 (0.0134) | | | |
| | 170 | Cont_PFCl_1 | | -0.31 (0.0153) | | | | |
| | 185 | Default_Temp_1 | | -0.30 (0.0177) | | | | |
| | 188 | Default_Temp_4 | | | | -0.31 (0.0128) | | |
| | 194 | Default_PFCm_4 | -0.37 (0.0032) | | | | | |
| | 199 | Default_PCC_2 | -0.29 (0.0195) | | | | | |

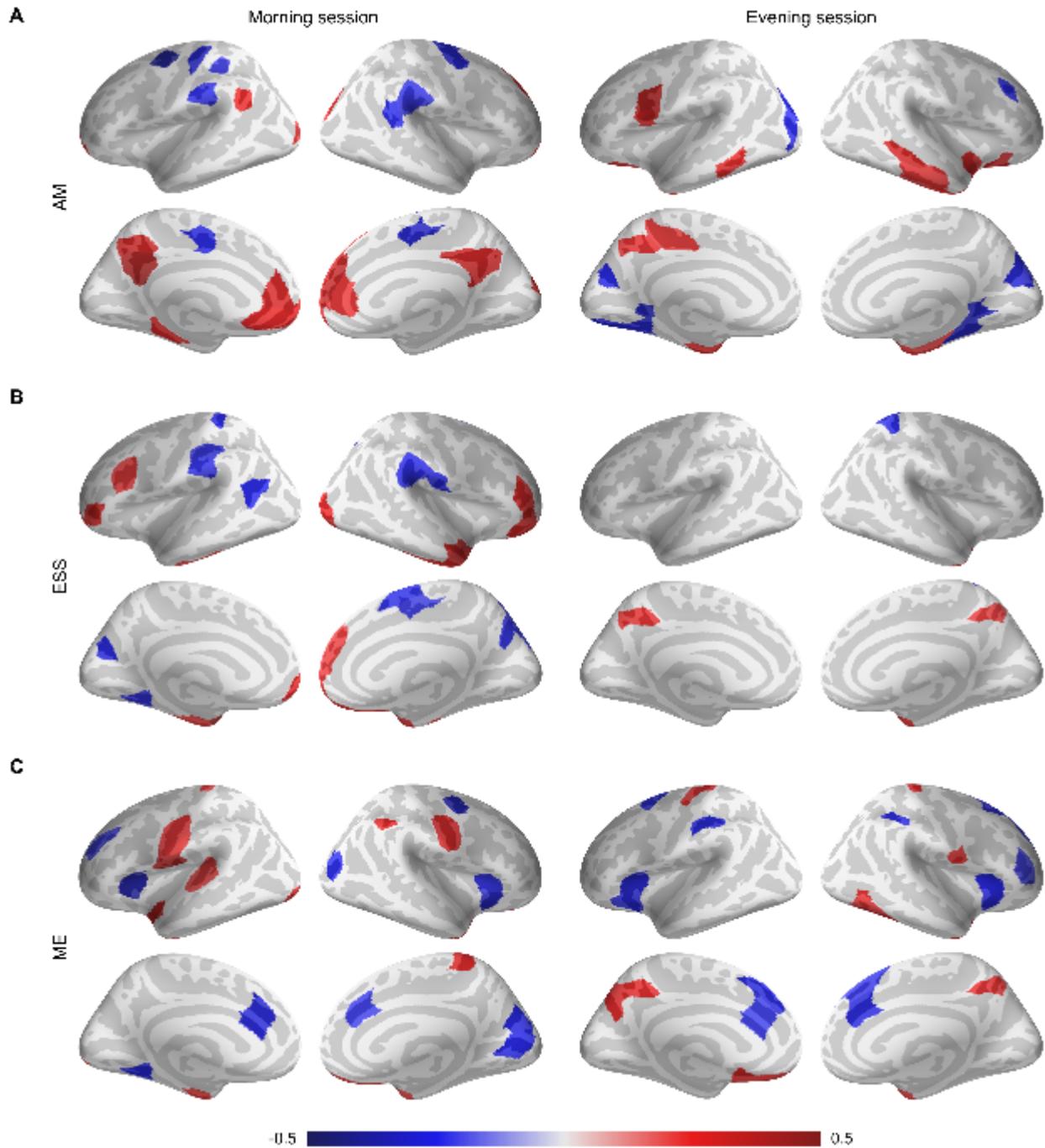

Figure 8. Correlation analysis between nodal centrality of brain regions and questionnaire variables AM (A), ESS (B), and ME (C). Nodes are colored according to their correlation magnitude.

## 4. Discussion

In this study, we used rs-fMRI and graph theory to examine diurnal fluctuation of whole-brain connectivity architectures across 62 healthy and young subjects, taking into account their chronotypes. The results of the study revealed meaningful information about the topological variations of the brain network during the day, as well as the associations of graph theoretical metrics with the variables of interest (e.g., ME, AM, and ESS scores), which are as follows: (1) among the global measures, there was a significant increase in small-worldness, modularity, assortativity, and network synchronization in the evening session compared to the morning session ($p < 0.05$, FDR corrected), however, there was no compelling evidence of changes in any of the global metrics in terms of chronotype (i.e., between morning- and evening-type participants); (2) local graph measures varied (during the day) predominantly across the somatomotor, attention, and default mode network ($p < 0.05$, FDR corrected), however, local metrics were mainly consistent with chronotype changes ($p > 0.05$); (3) analysis of the hubs demonstrated that the somatomotor network was the densest area of the brain (at both sessions but more during the evening) with both provincial and connector types, while hubs in the ventral attention network and visual network were primarily connector and provincial, respectively; (4) correlation analysis revealed significant associations between the variables derived from the questionnaires (such as ME, AM, and ESS) and the nodal characteristics of a number of brain regions in both sessions, most of which were related to the morning session.

### 4.1. Diurnal variations in global properties

A small-world network is an intermediary between random and regular networks, consisting a large number of short-range connections alongside a few long-range shortcuts (Watts and Strogatz, 1998). In the real system, there are other ways to define it, e.g., by considering the physical length of connections, all of which mathematically believe that small-world networks share a relative high transitivity and small mean geodesic (i.e., shortest-path) distance between nodes. This property makes them superior to other networks in terms of functional segregation (local specialization) and integration (global information flow), respectively (Bassett and Bullmore, 2006; Rubinov and Sporns, 2010). In our analysis of rs-fMRI data, we found high values of small-worldness for both scanning sessions ($\sigma \geq 1$ generally represents an efficient network), albeit with significant superiority in the evening session at sparse networks compared to the morning session, which reflects a more efficient functional topology in the brain network. Also, as another segregation metric, we found higher modularity in the evening session compared to the morning session at lower sparsity, reflecting a better division of the brain network into specialized communities. In another study, Anderson et al. (2016) explored how TOD affects functional brain networks in older healthy adults during rest and the performance of task. They found no topological changes during the resting-state, and a decrease in small-worldness and modularity in the afternoon (3:00 pm) compared to the morning (8:00 am), during the performance of a 1-back task.

More specifically, the hypothalamic suprachiasmatic nucleus (SCN) network, the control center for the body's "biological clock", is thought to be scale-free (Barabási and Albert, 1999; Hafner et al., 2012) and to possess small-world characteristics (Vasalou et al., 2009). In this regard, various studies have reported that more small-worldness of the SCN leads to more precise circadian fluctuations, a larger amplitude, higher synchrony and shifts more rapidly after the emergence of a new light/dark cycle compared to purely random or regular networks (Bodenstein et al., 2012; Gu et al., 2016; Hafner et al., 2012; Šimonka et al., 2015; Vasalou et al., 2009; Webb et al., 2012).

Furthermore, our findings revealed an increase in network assortativity over the course of the day, which overlaps with our previous results (Farahani et al., 2019). Increased assortativity is associated with a higher tendency for a node to link to other nodes with the same or similar degree (Newman, 2003; Foster et al., 2010), thus increasing the likelihood of a nearby hub supporting the faulty node. Finally, our results of network synchronization — a measure for assessing how well all nodes fluctuate in the same wave pattern — were not consistent with findings from Barahona and Pecora (2002), who presented that in networks of low redundancy, the small-world attitude results in more efficient synchrony than standard deterministic graphs, random graphs, and ideal constructive schemes. In fact, we found that in both the morning and evening sessions, as we move from low to high sparsity, small-worldness decreases, while synchronization increases. Notably, the network synchronization was significantly higher during the evening scanning session as compared to the morning. Overall, the findings of all global measures (i.e., small-worldness, modularity, assortativity, and synchronization) indicate that brain network organization varied throughout the day, which could be associated with increased brain function from morning to evening.

### 4.2. Nodal changes over the course of the day

To identify the nodal effects of TOD on the brain, a number of features, including degree centrality, betweenness centrality, clustering coeeficient, and nodal efficiency, were extracted from each of the 200 ROIs. Accordingly, the somatomotor network (e.g., the bilateral superior temporal, parietal, and postcentral gyri), attention network (e.g., left supramarginal, left middle frontal, right angular, and right superior parietal), as well as default mode network (e.g., the bilateral posterior cingulate cortex/precuneus and angular gyrus) experienced the most topological variations among brain networks during the day. The study of hubs in these networks also demonstared that they host several densely connected nodes, which strongly affect the brain functional integration and segregation.

Our resting-state findings showed that the within modular connectivity of the somatomotor network and its interaction with other parts of the brain (particularly, the ventral attention network) in both morning and evening sessions were relatively higher than other brain networks. Connectivity and hub analyses showed that the somatomotor network is the most densely-connected system in the brain predominantly during the evening session. We also found significantly increased FC in the bilateral postcentral, superior temporal, and superior parietal gyri as the waking time increases. Consistent with our results, such changes in the neural response across TOD have been previously reported in other rs-fMRI studies (Anderson et al., 2014; Fafrowicz et al., 2019; Jiang et al., 2016; Maire et al., 2018), as well as magnetoencephalography (MEG) studies to evaluate oscillatory activity at rest and during a finger-tapping task (Wilson et al., 2014). However, through a morphometric approach, conflicting results were found by Trefler et al. (2016) who discovered apparent decrease of cortical thickness as a function of TOD across the lateral surfaces of the left frontal, temporal, and parietal lobes. Together, our findings provide insight into how coordinated activity in sensorimotor networks vary over the course of the day. The increase of FC in somatomotor regions during the day indicate that neural synchronization is enhanced in these areas as a result of compensatory efforts to combat exhaustiveness, sleepiness, and absentmindedness after a day experience.

Ventral and dorsal attention networks regions are reported to be involved in stimulus-driven and goal-directed attention, respectively (Vossel et al., 2013). Connectivity and hub analyses showed that the ventral attention system is the second densely-connected network after the somatomotor areas mostly during the morning session with a focus on the frontal operculum insula and dorsal anterior cingulate gyrus. We also found a decreased FC throughout the course of the day within ventral areas such as the left supramarginal gyrus and left middle frontal gyrus (dorsal prefrontal cortex), in agreement with previous studies (Anderson et al., 2016; Jiang et al., 2016; Vandewalle et al., 2009). Also, changes were also observed in the right dorsal attention areas such as angular and superior parietal gyri. Drummond and Brown (2001) reported that total sleep deprivation was associated with increased activation in the prefrontal cortex and parietal lobes in learning and attention tasks, implying a compensatory mechanism for fighting against sleepiness. Overall, our results showed that due to increased sleepiness or fatigue during the day, the brain could not adapt to fulfill the demand for the attention and alertness even at early evening hours.

The DMN consists of a set of functionally connected and specialized neural units that contribute to many cognitive functions, including self-referential processing, autobiographical memory, and memory consolidation (Andrews-Hanna et al., 2007; Mayer et al., 2010; Raichle et al., 2001). Our study showed that the DMN is involved in relatively high neural activity at resting state during both sessions, particularly in the posterior cingulate cortex/precuneus which we detected it as a hub (higher in the evening than in the morning), and it plays a key role in the mediation of intrinsic activity through DMN (Fransson and Marrelec, 2008). Some areas within the DMN were found to be prone to variations in their connectivity profiles across daytime, such as the bilateral posterior cingulate cortex/precuneus, angular gyrus, superior temporal gyrus, and left superior frontal gyrus, a finding consistent with previous rs-fMRI studies (Facer-Childs et al., 2019b; Fransson, 2005; Jiang et al., 2016; Ku et al., 2018; Orban et al., 2020; Raichle et al., 2001; Shannon et al., 2013). Also, Blautzik et al. (2013) and Hodkinson et al. (2014) reported a rhythmic FC pattern of DMN during the day with its peak in the morning and declining trend in the afternoon. Diurnal changes in these areas indicate the functional coordination of the spatially disparate gyri (Jiang et al., 2016). Evidence has also pointed to a decrease in the default mode FC in sleep restriction as compared with the fully rested wakefulness (De Havas et al., 2012; Farahani et al., 2019). Considering these findings, total DMN neural activity decreased during the day, however, the FC increased in the hub regions such as posterior cingulate cortex/precuneus, which indicates the redoubled and regulatory efforts of these areas to balance of neuronal intercations (coupling or decoupling) within DMN subregions and adapt the network under continued wakeful condition across the course of a day.

### 4.3. Limitations and future directions

There are certain limitations associated with this study that should be considered in future research. First, brain nodes were derived from the cortical Schaefer/Yeo atlas (200-parcel/7-network parcellation; Schaefer et al., 2018). Schaefer's parcellations are available at multiple resolution (100 parcels to 1000 parcels), thus further investigation appraising network topology using finer parcellation schemes are warranted. Also, since the Schaefer atlas only considers cortical areas, whole-brain studies could be carried out by adding subcortical regions with the help of other atlases or segmentation algorithms. This issue should be investigated in more detail in future studies.

A second limitation concerns the measures used to compare the group-representative functional brain networks to one another. These measures tend to be correlated with one another — for example, a brain network with high efficiency necessarily must have shorter path length (Betzel et al., 2018). Thus, if one finds significant differences in one measure, it will probably be found in others. Therefore, given that we have chosen several correlated measures in this study, our analyses could be extended in future work to identify which of these measures is driving the others and how an exhaustive set of metrics can be designed to neurobiologically fit the study specification. Another issue about network measures is that global statistics are often non-specific (i.e., not especially informative). For example, the phrase "the patient group has lower efficiency than control" may not be clear to a neurosurgeon, so future work should be directed to investigate better interpretations of such metrics.

Yet another limitation concerns the applicability of small-world property. In real systems, early definitions of small-worldness, initiated by Watts and Strogatz (1998a), are ineffective for reasons such as confusing regular networks instead of small-world structure, neglecting the weight and physical length of connections, and lack of attention to the network density. Most of them present the network on the border of a circle, but real systems aren't embedded that way. Luckily, there are many different ways for a network to be small-world besides starting from a regular network and adding random links that reduce path length and researchers addressed the mentioned constraints by introducing several practical metrics (Bolaños et al., 2013; Muldoon et al., 2016; Rubinov and Sporns, 2010; Telesford et al., 2011). Applying these modified metrics in future work brings the study of small-world brain closer to reality.

Several theory-driven techniques have recently emerged to highlight the importance of machine learning, algorithmic optimization, and parallel computing in functional neuroimaging (Cohen et al., 2017; Douglas et al. 2013). For example, various algorithms called graph neural networks, of which Graph Convolutional Networks (GCN) being one of them, have been proposed that show how graph theory can be used as features to train deep learning models (Kipf and Welling, 2016; Wu et al., 2020) and discover neurological biomarkers using fMRI data (Li et al., 2020). As another example, a growing trend has developed in a family of algorithms called hyperalignment (or functional alignment) for projecting individuals' data into a common space based on how voxels respond to stimuli (Guntupalli et al., 2016; Haxby et al., 2011) or are connected to other voxels (Guntupalli et al., 2018; Haxby et al., 2020). The combined use of these techniques with network neuroscience will open a new generation of studies to transform knowledge of neural representations in complex brain networks.

## 5. Conclusion

Here, we present evidence for topological changes in functional brain network over the course of the day, i.e., in the morning as compared to the evening (1 and 10 hours after the wake-up time, respectively), using rs-fMRI data combined with graph theoretical analysis. Compared to the morning session, we found a more efficient functional topology in the evening session based on the global examination. Local properties differed prominently across the somatomotor, ventral attention, and default mode networks. In this way, we also considered inter-individual differences in diurnal preferences (chronotypes; early larks or night owls) in addition to the effect of TOD, however, we did not find any significant difference between the chronotypes. In summaty, our findings can reveal insight into diurnal variations in resting brain network and emphasize the role of TOD when designing neuroimaging experiments.

# Appendix

Table A1. Summary of Shaefer/Yeo parcellation's label name, component name, corresponding MNI coordinates, and the associated RGB code in the connectograms (7 networks, 200 nodes).

| | Label name | Component name | MNI coordinates | RGB code |
|---|---|---|---|---|
| Visual Network (VN), Left Hemisphere | | | | |
| 1 | LH_Vis_1 | Visual | -24,-53,-9 | 123,104,238 |
| 2 | LH_Vis_2 | Visual | -26,-77,-14 | 147,112,219 |
| 3 | LH_Vis_3 | Visual | -45,-69,-8 | 138,43,226 |
| 4 | LH_Vis_4 | Visual | -10,-67,-4 | 201,160,220 |
| 5 | LH_Vis_5 | Visual | -27,-95,-12 | 204,204,255 |
| 6 | LH_Vis_6 | Visual | -14,-44,-3 | 75,0,130 |
| 7 | LH_Vis_7 | Visual | -5,-93,-4 | 181,126,220 |
| 8 | LH_Vis_8 | Visual | -47,-70,10 | 147,112,219 |
| 9 | LH_Vis_9 | Visual | -23,-97,6 | 167,107,207 |
| 10 | LH_Vis_10 | Visual | -11,-70,7 | 116,108,192 |
| 11 | LH_Vis_11 | Visual | -40,-85,11 | 138,43,226 |
| 12 | LH_Vis_12 | Visual | -12,-73,22 | 111,0,255 |
| 13 | LH_Vis_13 | Visual | -7,-87,28 | 120,81,169 |
| 14 | LH_Vis_14 | Visual | -23,-87,23 | 115,79,150 |
| Somatomotor Network (SMN), Left Hemisphere | | | | |
| 15 | LH_SomMot_1 | Somatomotor | -51,-4,-2 | 135,206,235 |
| 16 | LH_SomMot_2 | Somatomotor | -53,-24,9 | 30,144,255 |
| 17 | LH_SomMot_3 | Somatomotor | -37,-21,16 | 0,191,255 |
| 18 | LH_SomMot_4 | Somatomotor | -55,-4,10 | 0,0,205 |
| 19 | LH_SomMot_5 | Somatomotor | -53,-22,18 | 172,229,238 |
| 20 | LH_SomMot_6 | Somatomotor | -56,-8,31 | 135,206,250 |
| 21 | LH_SomMot_7 | Somatomotor | -47,-9,46 | 119,181,254 |
| 22 | LH_SomMot_8 | Somatomotor | -7,-12,46 | 79,134,247 |
| 23 | LH_SomMot_9 | Somatomotor | -49,-28,57 | 119,158,203 |
| 24 | LH_SomMot_10 | Somatomotor | -40,-25,57 | 65,102,245 |
| 25 | LH_SomMot_11 | Somatomotor | -31,-46,63 | 69,177,232 |
| 26 | LH_SomMot_12 | Somatomotor | -32,-22,64 | 49,140,231 |
| 27 | LH_SomMot_13 | Somatomotor | -26,-38,68 | 73,151,208 |
| 28 | LH_SomMot_14 | Somatomotor | -20,-11,68 | 15,192,252 |
| 29 | LH_SomMot_15 | Somatomotor | -5,-29,67 | 65,125,193 |
| 30 | LH_SomMot_16 | Somatomotor | -19,-31,68 | 0,127,255 |
| Dorsal Attention Network (DAN), Left Hemisphere | | | | |
| 31 | LH_DorsAttn_Post_1 | Posterior | -43,-48,-19 | 0,255,127 |
| 32 | LH_DorsAttn_Post_2 | Posterior | -57,-60,-1 | 50,205,50 |
| 33 | LH_DorsAttn_Post_3 | Posterior | -26,-70,38 | 173,255,47 |
| 34 | LH_DorsAttn_Post_4 | Posterior | -54,-27,42 | 144,238,144 |
| 35 | LH_DorsAttn_Post_5 | Posterior | -41,-35,47 | 60,179,113 |
| 36 | LH_DorsAttn_Post_6 | Posterior | -33,-49,47 | 34,139,34 |
| 37 | LH_DorsAttn_Post_7 | Posterior | -17,-73,54 | 152,255,152 |
| 38 | LH_DorsAttn_Post_8 | Posterior | -29,-60,59 | 144,238,144 |
| 39 | LH_DorsAttn_Post_9 | Posterior | -6,-60,57 | 119,221,119 |
| 40 | LH_DorsAttn_Post_10 | Posterior | -17,-53,68 | 116,195,101 |
| 41 | LH_DorsAttn_FEF_1 | Frontal Eye Fields | -31,-4,53 | 80,200,120 |
| 42 | LH_DorsAttn_FEF_2 | Frontal Eye Fields | -22,6,62 | 57,255,20 |
| 43 | LH_DorsAttn_PrCv_1 | Precentral Ventral | -48,6,29 | 34,139,34 |
| Ventral Attention Network (DAN), Left Hemisphere | | | | |
| 44 | LH_SalVentAttn_ParOper_1 | Parietal Operculum | -56,-40,20 | 249,132,229 |
| 45 | LH_SalVentAttn_ParOper_2 | Parietal Operculum | -61,-26,28 | 254,78,218 |
| 46 | LH_SalVentAttn_ParOper_3 | Parietal Operculum | -60,-39,36 | 207,113,175 |
| 47 | LH_SalVentAttn_FrOperIns_1 | Frontal Operculum Insula | -39,-4,-4 | 189,51,164 |
| 48 | LH_SalVentAttn_FrOperIns_2 | Frontal Operculum Insula | -33,20,5 | 204,0,204 |
| 49 | LH_SalVentAttn_FrOperIns_3 | Frontal Operculum Insula | -39,1,11 | 218,112,214 |
| 50 | LH_SalVentAttn_FrOperIns_4 | Frontal Operculum Insula | -51,9,11 | 241,167,254 |
| 51 | LH_SalVentAttn_PFCl_1 | Lateral Prefrontal Cortex | -28,43,31 | 238,130,238 |
| 52 | LH_SalVentAttn_Med_1 | Medial | -6,9,41 | 255,111,255 |
| 53 | LH_SalVentAttn_Med_2 | Medial | -11,-35,46 | 207,52,118 |
| 54 | LH_SalVentAttn_Med_3 | Medial | -6,-3,65 | 223,0,255 |
| Limbic Network (LN), Left Hemisphere | | | | |
| 55 | LH_Limbic_OFC_1 | Orbital Frontal Cortex | -24,22,-20 | 255,248,220 |
| 56 | LH_Limbic_OFC_2 | Orbital Frontal Cortex | -10,35,-21 | 240,230,140 |
| 57 | LH_Limbic_TempPole_1 | Temporal Pole | -29,-6,-39 | 252,247,94 |
| 58 | LH_Limbic_TempPole_2 | Temporal Pole | -45,-20,-30 | 255,250,205 |

| # | Label | Region | Coords | RGB |
|---|---|---|---|---|
| 59 | LH_Limbic_TempPole_3 | Temporal Pole | -28,10,-34 | 251,236,93 |
| 60 | LH_Limbic_TempPole_4 | Temporal Pole | -43,8,-19 | 255,247,0 |
| Frontoparietal Network (FPN), Left Hemisphere | | | | |
| 61 | LH_Cont_Par_1 | Parietal | -53,-51,46 | 255,165,0 |
| 62 | LH_Cont_Par_2 | Parietal | -35,-62,48 | 255,140,0 |
| 63 | LH_Cont_Par_3 | Parietal | -45,-42,46 | 255,160,137 |
| 64 | LH_Cont_Temp_1 | Temporal | -61,-43,-13 | 255,200,124 |
| 65 | LH_Cont_OFC_1 | Orbital Frontal Cortex | -32,42,-13 | 255,153,102 |
| 66 | LH_Cont_PFCl_1 | Lateral Prefrontal Cortex | -42,49,-6 | 255,163,67 |
| 67 | LH_Cont_PFCl_2 | Lateral Prefrontal Cortex | -28,58,8 | 255,130,67 |
| 68 | LH_Cont_PFCl_3 | Lateral Prefrontal Cortex | -42,40,16 | 255,174,66 |
| 69 | LH_Cont_PFCl_4 | Lateral Prefrontal Cortex | -44,20,27 | 237,135,45 |
| 70 | LH_Cont_PFCl_5 | Lateral Prefrontal Cortex | -43,6,43 | 224,141,60 |
| 71 | LH_Cont_pCun_1 | Precuneus | -9,-73,38 | 255,153,51 |
| 72 | LH_Cont_Cing_1 | Cingulate | -5,-29,28 | 237,145,33 |
| 73 | LH_Cont_Cing_2 | Cingulate | -3,4,30 | 251,153,2 |
| Default Mode Network (DMN), Left Hemisphere | | | | |
| 74 | LH_Default_Temp_1 | Temporal | -47,8,-33 | 240,128,128 |
| 75 | LH_Default_Temp_2 | Temporal | -60,-19,-22 | 255,69,0 |
| 76 | LH_Default_Temp_3 | Temporal | -56,-6,-12 | 165,42,42 |
| 77 | LH_Default_Temp_4 | Temporal | -58,-30,-4 | 255,0,0 |
| 78 | LH_Default_Temp_5 | Temporal | -58,-43,7 | 123,17,19 |
| 79 | LH_Default_Par_1 | Parietal | -48,-57,18 | 204,51,51 |
| 80 | LH_Default_Par_2 | Parietal | -39,-80,31 | 205,92,92 |
| 81 | LH_Default_Par_3 | Parietal | -57,-54,28 | 253,94,83 |
| 82 | LH_Default_Par_4 | Parietal | -46,-66,38 | 127,23,52 |
| 83 | LH_Default_PFC_1 | Prefrontal Cortex | -35,20,-13 | 255,53,94 |
| 84 | LH_Default_PFC_2 | Prefrontal Cortex | -6,36,-10 | 235,76,66 |
| 85 | LH_Default_PFC_3 | Prefrontal Cortex | -46,31,-7 | 204,78,92 |
| 86 | LH_Default_PFC_4 | Prefrontal Cortex | -12,63,-6 | 178,34,34 |
| 87 | LH_Default_PFC_5 | Prefrontal Cortex | -52,22,8 | 203,65,84 |
| 88 | LH_Default_PFC_6 | Prefrontal Cortex | -6,44,7 | 237,28,36 |
| 89 | LH_Default_PFC_7 | Prefrontal Cortex | -8,59,21 | 218,44,67 |
| 90 | LH_Default_PFC_8 | Prefrontal Cortex | -6,30,25 | 229,26,76 |
| 91 | LH_Default_PFC_9 | Prefrontal Cortex | -11,47,45 | 255,36,0 |
| 92 | LH_Default_PFC_10 | Prefrontal Cortex | -3,33,43 | 255,69,0 |
| 93 | LH_Default_PFC_11 | Prefrontal Cortex | -40,19,49 | 171,75,82 |
| 94 | LH_Default_PFC_12 | Prefrontal Cortex | -24,25,49 | 156,37,66 |
| 95 | LH_Default_PFC_13 | Prefrontal Cortex | -9,17,63 | 194,59,34 |
| 96 | LH_Default_pCunPCC_1 | Precuneus/Posterior Cingulate Cortex | -11,-56,13 | 196,30,58 |
| 97 | LH_Default_pCunPCC_2 | Precuneus/Posterior Cingulate Cortex | -5,-55,27 | 255,64,64 |
| 98 | LH_Default_pCunPCC_3 | Precuneus/Posterior Cingulate Cortex | -4,-31,36 | 211,0,63 |
| 99 | LH_Default_pCunPCC_4 | Precuneus/Posterior Cingulate Cortex | -6,-54,42 | 157,41,51 |
| 100 | LH_Default_PHC_1 | Parahippocampal Cortex | -26,-32,-18 | 205,92,92 |
| Visual Network (VN), Right Hemisphere | | | | |
| 101 | RH_Vis_1 | Visual | 39,-35,-23 | 123,104,238 |
| 102 | RH_Vis_2 | Visual | 28,-36,-14 | 147,112,219 |
| 103 | RH_Vis_3 | Visual | 29,-69,-12 | 138,43,226 |
| 104 | RH_Vis_4 | Visual | 12,-65,-5 | 201,160,220 |
| 105 | RH_Vis_5 | Visual | 48,-71,-6 | 204,204,255 |
| 106 | RH_Vis_6 | Visual | 11,-92,-5 | 75,0,130 |
| 107 | RH_Vis_7 | Visual | 16,-46,-1 | 181,126,220 |
| 108 | RH_Vis_8 | Visual | 31,-94,-4 | 147,112,219 |
| 109 | RH_Vis_9 | Visual | 9,-75,9 | 167,107,207 |
| 110 | RH_Vis_10 | Visual | 22,-60,7 | 116,108,192 |
| 111 | RH_Vis_11 | Visual | 42,-80,10 | 138,43,226 |
| 112 | RH_Vis_12 | Visual | 20,-90,22 | 111,0,255 |
| 113 | RH_Vis_13 | Visual | 11,-74,26 | 120,81,169 |
| 114 | RH_Vis_14 | Visual | 16,-85,39 | 115,79,150 |
| 115 | RH_Vis_15 | Visual | 33,-75,32 | 105,53,156 |
| Somatomotor Network (SMN), Right Hemisphere | | | | |
| 116 | RH_SomMot_1 | Somatomotor | 51,-15,5 | 135,206,235 |
| 117 | RH_SomMot_2 | Somatomotor | 64,-23,8 | 30,144,255 |
| 118 | RH_SomMot_3 | Somatomotor | 38,-13,15 | 0,191,255 |
| 119 | RH_SomMot_4 | Somatomotor | 44,-27,18 | 0,0,205 |
| 120 | RH_SomMot_5 | Somatomotor | 59,0,10 | 172,229,238 |
| 121 | RH_SomMot_6 | Somatomotor | 56,-11,14 | 135,206,250 |
| 122 | RH_SomMot_7 | Somatomotor | 58,-5,31 | 119,181,254 |
| 123 | RH_SomMot_8 | Somatomotor | 10,-15,41 | 79,134,247 |
| 124 | RH_SomMot_9 | Somatomotor | 51,-22,52 | 119,158,203 |
| 125 | RH_SomMot_10 | Somatomotor | 47,-11,48 | 65,102,245 |
| 126 | RH_SomMot_11 | Somatomotor | 7,-11,51 | 69,177,232 |
| 127 | RH_SomMot_12 | Somatomotor | 40,-24,57 | 49,140,231 |
| 128 | RH_SomMot_13 | Somatomotor | 32,-40,64 | 73,151,208 |
| 129 | RH_SomMot_14 | Somatomotor | 33,-21,65 | 15,192,252 |
| 130 | RH_SomMot_15 | Somatomotor | 29,-34,65 | 65,125,193 |
| 131 | RH_SomMot_16 | Somatomotor | 22,-9,67 | 0,127,255 |
| 132 | RH_SomMot_17 | Somatomotor | 10,-39,69 | 0,191,255 |
| 133 | RH_SomMot_18 | Somatomotor | 6,-23,69 | 29,172,214 |
| 134 | RH_SomMot_19 | Somatomotor | 20,-29,70 | 25,116,210 |
| Dorsal Attention Network (DAN), Right Hemisphere | | | | |
| 135 | RH_DorsAttn_Post_1 | Posterior | 50,-53,-15 | 0,255,127 |
| 136 | RH_DorsAttn_Post_2 | Posterior | 52,-60,9 | 50,205,50 |
| 137 | RH_DorsAttn_Post_3 | Posterior | 59,-16,34 | 173,255,47 |
| 138 | RH_DorsAttn_Post_4 | Posterior | 46,-38,49 | 144,238,144 |
| 139 | RH_DorsAttn_Post_5 | Posterior | 41,-31,46 | 60,179,113 |
| 140 | RH_DorsAttn_Post_6 | Posterior | 15,-73,53 | 34,139,34 |
| 141 | RH_DorsAttn_Post_7 | Posterior | 34,-48,51 | 152,255,152 |
| 142 | RH_DorsAttn_Post_8 | Posterior | 26,-61,58 | 144,238,144 |
| 143 | RH_DorsAttn_Post_9 | Posterior | 8,-56,61 | 119,221,119 |
| 144 | RH_DorsAttn_Post_10 | Posterior | 21,-48,70 | 116,195,101 |
| 145 | RH_DorsAttn_FEF_1 | Frontal Eye Fields | 34,-4,52 | 80,200,120 |
| 146 | RH_DorsAttn_FEF_2 | Frontal Eye Fields | 26,7,58 | 57,255,20 |
| 147 | RH_DorsAttn_PrCv_1 | Precentral Ventral | 52,11,21 | 34,139,34 |
| Ventral Attention Network (VAN), Right Hemisphere | | | | |
| 148 | RH_SalVentAttn_TempOccPar_1 | Temporal Occipital Parietal | 57,-45,9 | 255,0,144 |
| 149 | RH_SalVentAttn_TempOccParr_2 | Temporal Occipital Parietal | 60,-39,17 | 218,29,129 |
| 150 | RH_SalVentAttn_TempOccParr_3 | Temporal Occipital Parietal | 60,-26,27 | 255,111,255 |
| 151 | RH_SalVentAttn_PrC_1 | Precentral | 51,4,40 | 189,51,164 |
| 152 | RH_SalVentAttn_FrOperIns_1 | Frontal Operculum Insula | 41,6,-15 | 189,51,164 |
| 153 | RH_SalVentAttn_FrOperIns_2 | Frontal Operculum Insula | 46,-4,-4 | 204,0,204 |
| 154 | RH_SalVentAttn_FrOperIns_3 | Frontal Operculum Insula | 36,24,5 | 218,112,214 |
| 155 | RH_SalVentAttn_FrOperIns_4 | Frontal Operculum Insula | 43,7,4 | 241,167,254 |
| 156 | RH_SalVentAttn_Med_1 | Medial | 7,9,41 | 255,111,255 |
| 157 | RH_SalVentAttn_Med_2 | Medial | 11,-36,47 | 207,52,118 |
| 158 | RH_SalVentAttn_Med_3 | Medial | 8,3,66 | 223,0,255 |
| Limbic Network (LN), Right Hemisphere | | | | |
| 159 | RH_Limbic_OFC_1 | Orbital Frontal Cortex | 12,39,-22 | 255,248,220 |
| 160 | RH_Limbic_OFC_2 | Orbital Frontal Cortex | 28,22,-19 | 240,230,140 |
| 161 | RH_Limbic_OFC_3 | Orbital Frontal Cortex | 15,64,-8 | 253,253,150 |
| 162 | RH_Limbic_TempPole_1 | Temporal Pole | 30,9,-38 | 252,247,94 |
| 163 | RH_Limbic_TempPole_2 | Temporal Pole | 47,-12,-35 | 255,250,205 |
| 164 | RH_Limbic_TempPole_3 | Temporal Pole | 25,-11,-32 | 251,236,93 |
| Frontoparietal Network (FPN), Right Hemisphere | | | | |
| 165 | RH_Cont_Par_1 | Parietal | 62,-37,37 | 255,165,0 |
| 166 | RH_Cont_Par_2 | Parietal | 53,-42,48 | 255,140,0 |
| 167 | RH_Cont_Par_3 | Parietal | 37,-63,47 | 255,160,137 |
| 168 | RH_Cont_Temp_1 | Temporal | 63,-41,-12 | 255,200,124 |
| 169 | RH_Cont_PFCv_1 | Ventral Prefrontal Cortex | 34,21,-8 | 255,103,0 |
| 170 | RH_Cont_PFCl_1 | Lateral Prefrontal Cortex | 36,46,-13 | 255,153,102 |
| 171 | RH_Cont_PFCl_2 | Lateral Prefrontal Cortex | 29,58,5 | 255,163,67 |
| 172 | RH_Cont_PFCl_3 | Lateral Prefrontal Cortex | 43,45,10 | 255,130,67 |
| 173 | RH_Cont_PFCl_4 | Lateral Prefrontal Cortex | 46,24,26 | 255,174,66 |
| 174 | RH_Cont_PFCl_5 | Lateral Prefrontal Cortex | 30,48,27 | 237,135,45 |
| 175 | RH_Cont_PFCl_6 | Lateral Prefrontal Cortex | 41,33,37 | 224,141,60 |
| 176 | RH_Cont_PFCl_7 | Lateral Prefrontal Cortex | 42,14,49 | 255,167,0 |
| 177 | RH_Cont_pCun_1 | Precuneus | 14,-70,37 | 255,153,51 |
| 178 | RH_Cont_Cing_1 | Cingulate | 5,-24,31 | 237,145,33 |
| 179 | RH_Cont_Cing_2 | Cingulate | 5,3,30 | 251,153,2 |
| 180 | RH_Cont_PFCmp_1 | Medial Posterior Prefrontal Cortex | 7,31,28 | 255,186,0 |
| 181 | RH_Cont_PFCmp_2 | Medial Posterior Prefrontal Cortex | 7,25,55 | 228,155,15 |
| Default Mode Network (DMN), Right Hemisphere | | | | |
| 182 | RH_Default_Par_1 | Parietal | 47,-69,27 | 153,0,0 |
| 183 | RH_Default_Par_2 | Parietal | 54,-50,28 | 228,113,122 |
| 184 | RH_Default_Par_3 | Parietal | 51,-59,44 | 234,60,83 |
| 185 | RH_Default_Temp_1 | Temporal | 47,13,-30 | 240,128,128 |
| 186 | RH_Default_Temp_2 | Temporal | 61,-13,-21 | 255,69,0 |
| 187 | RH_Default_Temp_3 | Temporal | 55,-6,-10 | 165,42,42 |
| 188 | RH_Default_Temp_4 | Temporal | 63,-27,-6 | 255,0,0 |
| 189 | RH_Default_Temp_5 | Temporal | 52,-31,2 | 123,17,19 |
| 190 | RH_Default_PFCv_1 | Ventral Prefrontal Cortex | 51,28,0 | 255,64,64 |
| 191 | RH_Default_PFCd/m_1 | Dorsal/Medial Prefrontal Cortex | 5,37,-14 | 220,20,60 |
| 192 | RH_Default_PFCd/m_2 | Dorsal/Medial Prefrontal Cortex | 8,42,4 | 227,66,52 |
| 193 | RH_Default_PFCd/m_3 | Dorsal/Medial Prefrontal Cortex | 6,29,15 | 215,59,62 |
| 194 | RH_Default_PFCd/m_4 | Dorsal/Medial Prefrontal Cortex | 8,58,18 | 203,65,84 |
| 195 | RH_Default_PFCd/m_5 | Dorsal/Medial Prefrontal Cortex | 15,46,44 | 255,83,73 |
| 196 | RH_Default_PFCd/m_6 | Dorsal/Medial Prefrontal Cortex | 29,30,42 | 206,32,41 |
| 197 | RH_Default_PFCd/m_7 | Dorsal/Medial Prefrontal Cortex | 23,24,53 | 232,0,13 |
| 198 | RH_Default_pCunPCC_1 | Precuneus/Posterior Cingulate Cortex | 12,-55,15 | 196,30,58 |
| 199 | RH_Default_pCunPCC_2 | Precuneus/Posterior Cingulate Cortex | 7,-49,31 | 255,64,64 |
| 200 | RH_Default_pCunPCC_3 | Precuneus/Posterior Cingulate Cortex | 6,-58,44 | 211,0,63 |